\documentclass[aps,preprint,tightenlines,superscriptaddress,showpacs,byrevtex]{revtex4}
\usepackage{graphicx}
\usepackage{epsfig}
\usepackage{dcolumn}
\usepackage{bm}
\usepackage{color}
\usepackage{amsmath}
\newcommand{\gisr}{\gamma_{\rm ISR}}
\newcommand{\gev}{\rm GeV}
\newcommand{\gevc}{{\rm GeV}/c}
\newcommand{\gevcs}{{\rm GeV}/c^2}
\newcommand{\mev}{\rm MeV}
\newcommand{\mevcs}{{\rm MeV}/c^2}

\newcommand{\ev}{\rm eV}

\newcommand{\infb}{\rm fb^{-1}}
\newcommand{\MMS}{M_{\rm rec}^2}
\newcommand{\fz}{f_0(980)}

\newcommand{\lum}{{\cal L}}
\newcommand{\eff}{\varepsilon}
\newcommand{\BR}{{\cal B}}
\newcommand{\jpc}{J^{PC}}
\newcommand{\pip}{\pi^+}
\newcommand{\pim}{\pi^-}

\newcommand{\psp}{\psi(2S)}

\newcommand{\jpsi}{J/\psi}
\newcommand{\psift}{\psi(4040)}
\newcommand{\psifto}{\psi(4160)}

\newcommand{\EE}{e^+e^-}
\newcommand{\MM}{\mu^+\mu^-}
\newcommand{\LL}{\ell^+\ell^-}

\newcommand{\pp}{\pi^+\pi^-}

\newcommand{\ccb}{c\bar{c}}

\newcommand{\ppjpsi}{\pi^+\pi^- J/\psi}

\newcommand{\reduline}{\bgroup\markoverwith
{\textcolor{red}{\rule[0.5ex]{2pt}{0.4pt}}}\ULon}

\newcommand{\beq}{\begin{equation}}
\newcommand{\eeq}{\end{equation}}
\newcommand{\bitm}{\begin{itemize}}
\newcommand{\eitm}{\end{itemize}}

\def\Journal#1#2#3#4{{#1} {\bf #2}, #3 (#4)}
\def\IJMP{Int. J. Mod. Phys. A}

\def\NIMA{Nucl. Instrum. Methods A}

\def\PRL{Phys. Rev. Lett.}
\def\PRD{Phys. Rev. D}

\def\EPJC{Eur. Phys. J. C}

\def\PTEP{Prog. Theor. Exp. Phys.}

\begin{document}

\preprint{} \preprint{ \vbox{ \hbox{   }
                        \hbox{Belle Preprint 2014-17}
                        \hbox{KEK   Preprint 2014-29}
        }}
\title{
\quad\\[2.0cm]
Measurement of $\EE\to \pp\psp$ via Initial State Radiation at Belle}

\noaffiliation
\affiliation{University of the Basque Country UPV/EHU, 48080 Bilbao}
\affiliation{Beihang University, Beijing 100191}
\affiliation{Budker Institute of Nuclear Physics SB RAS and Novosibirsk State University, Novosibirsk 630090}
\affiliation{Faculty of Mathematics and Physics, Charles University, 121 16 Prague}
\affiliation{University of Cincinnati, Cincinnati, Ohio 45221}
\affiliation{Deutsches Elektronen--Synchrotron, 22607 Hamburg}
\affiliation{Justus-Liebig-Universit\"at Gie\ss{}en, 35392 Gie\ss{}en}
\affiliation{The Graduate University for Advanced Studies, Hayama 240-0193}
\affiliation{Gyeongsang National University, Chinju 660-701}
\affiliation{Hanyang University, Seoul 133-791}
\affiliation{University of Hawaii, Honolulu, Hawaii 96822}
\affiliation{High Energy Accelerator Research Organization (KEK), Tsukuba 305-0801}
\affiliation{IKERBASQUE, Basque Foundation for Science, 48011 Bilbao}
\affiliation{Indian Institute of Technology Guwahati, Assam 781039}
\affiliation{Indian Institute of Technology Madras, Chennai 600036}
\affiliation{Institute of High Energy Physics, Chinese Academy of Sciences, Beijing 100049}
\affiliation{Institute of High Energy Physics, Vienna 1050}
\affiliation{Institute for High Energy Physics, Protvino 142281}
\affiliation{INFN - Sezione di Torino, 10125 Torino}
\affiliation{Institute for Theoretical and Experimental Physics, Moscow 117218}
\affiliation{J. Stefan Institute, 1000 Ljubljana}
\affiliation{Kanagawa University, Yokohama 221-8686}
\affiliation{Kennesaw State University, Kennesaw GA 30144}
\affiliation{Department of Physics, Faculty of Science, King Abdulaziz University, Jeddah 21589}
\affiliation{Korea Institute of Science and Technology Information, Daejeon 305-806}
\affiliation{Korea University, Seoul 136-713}
\affiliation{Kyungpook National University, Daegu 702-701}
\affiliation{\'Ecole Polytechnique F\'ed\'erale de Lausanne (EPFL), Lausanne 1015}
\affiliation{Luther College, Decorah, Iowa 52101}
\affiliation{University of Maribor, 2000 Maribor}
\affiliation{Max-Planck-Institut f\"ur Physik, 80805 M\"unchen}
\affiliation{School of Physics, University of Melbourne, Victoria 3010}
\affiliation{Moscow Physical Engineering Institute, Moscow 115409}
\affiliation{Moscow Institute of Physics and Technology, Moscow Region 141700}
\affiliation{Graduate School of Science, Nagoya University, Nagoya 464-8602}
\affiliation{Kobayashi-Maskawa Institute, Nagoya University, Nagoya 464-8602}
\affiliation{Nara Women's University, Nara 630-8506}
\affiliation{National Central University, Chung-li 32054}
\affiliation{Department of Physics, National Taiwan University, Taipei 10617}
\affiliation{H. Niewodniczanski Institute of Nuclear Physics, Krakow 31-342}
\affiliation{Niigata University, Niigata 950-2181}
\affiliation{Osaka City University, Osaka 558-8585}
\affiliation{Pacific Northwest National Laboratory, Richland, Washington 99352}
\affiliation{Peking University, Beijing 100871}
\affiliation{University of Pittsburgh, Pittsburgh, Pennsylvania 15260}
\affiliation{University of Science and Technology of China, Hefei 230026}
\affiliation{Seoul National University, Seoul 151-742}
\affiliation{Soongsil University, Seoul 156-743}
\affiliation{Sungkyunkwan University, Suwon 440-746}
\affiliation{School of Physics, University of Sydney, NSW 2006}
\affiliation{Department of Physics, Faculty of Science, University of Tabuk, Tabuk 71451}
\affiliation{Tata Institute of Fundamental Research, Mumbai 400005}
\affiliation{Excellence Cluster Universe, Technische Universit\"at M\"unchen, 85748 Garching}
\affiliation{Toho University, Funabashi 274-8510}
\affiliation{Tohoku University, Sendai 980-8578}
\affiliation{Department of Physics, University of Tokyo, Tokyo 113-0033}
\affiliation{Tokyo Institute of Technology, Tokyo 152-8550}
\affiliation{Tokyo Metropolitan University, Tokyo 192-0397}
\affiliation{University of Torino, 10124 Torino}
\affiliation{CNP, Virginia Polytechnic Institute and State University, Blacksburg, Virginia 24061}
\affiliation{Wayne State University, Detroit, Michigan 48202}
\affiliation{Yamagata University, Yamagata 990-8560}
\affiliation{Yonsei University, Seoul 120-749}
  \author{X.~L.~Wang}\affiliation{CNP, Virginia Polytechnic Institute and State University, Blacksburg, Virginia 24061} 
  \author{C.~Z.~Yuan}\affiliation{Institute of High Energy Physics, Chinese Academy of Sciences, Beijing 100049} 
  \author{C.~P.~Shen}\affiliation{Beihang University, Beijing 100191} 
  \author{P.~Wang}\affiliation{Institute of High Energy Physics, Chinese Academy of Sciences, Beijing 100049} 
  \author{A.~Abdesselam}\affiliation{Department of Physics, Faculty of Science, University of Tabuk, Tabuk 71451} 
  \author{I.~Adachi}\affiliation{High Energy Accelerator Research Organization (KEK), Tsukuba 305-0801}\affiliation{The Graduate University for Advanced Studies, Hayama 240-0193} 
  \author{H.~Aihara}\affiliation{Department of Physics, University of Tokyo, Tokyo 113-0033} 
  \author{S.~Al~Said}\affiliation{Department of Physics, Faculty of Science, University of Tabuk, Tabuk 71451}\affiliation{Department of Physics, Faculty of Science, King Abdulaziz University, Jeddah 21589} 
  \author{K.~Arinstein}\affiliation{Budker Institute of Nuclear Physics SB RAS and Novosibirsk State University, Novosibirsk 630090} 
  \author{D.~M.~Asner}\affiliation{Pacific Northwest National Laboratory, Richland, Washington 99352} 
% ...
\collaboration{The Belle Collaboration}

\date{\today}

\begin{abstract}

We report measurement of the cross section of $\EE\to \pp\psp$ between 4.0 and
5.5~$\gev$, based on an analysis of initial state radiation events in a  $980~\infb$
 data sample recorded with the Belle detector.
The properties of the $Y(4360)$ and $Y(4660)$ states are
determined. Fitting the mass spectrum of $\pp\psp$ with two
coherent Breit-Wigner functions, we find two solutions with
identical mass and width but different couplings to
electron-positron pairs: $M_{Y(4360)} = (4347\pm 6\pm 3)~\mevcs$,
$\Gamma_{Y(4360)} = (103\pm 9\pm 5)~\mev$, $M_{Y(4660)} =
(4652\pm10\pm 8)~\mevcs$, $\Gamma_{Y(4660)} = (68\pm 11\pm 1)~\mev$;
and $\BR[Y(4360)\to \pp\psp]\cdot \Gamma_{Y(4360)}^{\EE} = (10.9\pm
0.6\pm 0.7)~\ev$ and $\BR[Y(4660)\to \pp\psp]\cdot
\Gamma_{Y(4660)}^{\EE} = (8.1\pm 1.1\pm 0.5)~\ev$ for one solution;
or $\BR[Y(4360)\to \pp\psp]\cdot \Gamma_{Y(4360)}^{\EE} = (9.2\pm
0.6\pm 0.6)~\ev$ and $\BR[Y(4660)\to \pp\psp]\cdot
\Gamma_{Y(4660)}^{\EE} = (2.0\pm 0.3\pm 0.2)~\ev$ for the other.
Here, the first errors are statistical and the second systematic.
Evidence for a charged charmoniumlike structure at $4.05~\gevcs$ is
observed in the $\pi^{\pm}\psp$ intermediate state in the $Y(4360)$
decays.

\end{abstract}


\maketitle

\section{Introduction}
\label{part1}

Many charmonium and charmoniumlike states have been discovered in
the past decade. Some are good candidates for conventional
charmonium states, while others exhibit unusual properties
consistent with expectations for exotic states such as
tetraquarks, molecules, hybrids, hadrocharmonia, or
glueballs~\cite{review,ycz_review}. The initial state radiation
(ISR) technique has played a very important role in the discovery
and studies of a number of the charmonium and charmoniumlike states.
The $\jpc$ quantum numbers of the final states accompanying
the ISR photon(s) are restricted to $\jpc = 1^{--}$ and so
favors this technique for the study of vector particles.

The BaBar experiment observed the $Y(4260)$ state in the
process $\EE\to \gamma_{\rm ISR} \pi^+\pi^-J/\psi$~\cite{babay4260},
and this was confirmed by the CLEO~\cite{cleoy} and Belle
experiments~\cite{belley} with the same technique. Moreover, Belle
reported a broad structure near 4.0~$\gev$ that they dubbed the
$Y(4008)$. In an analysis of the $\EE\to \gamma_{\rm ISR}
\pi^+\pi^-\psi(2S)$ process, BaBar found a structure near
4.32~$\gev$~\cite{babay4324}, while Belle observed two resonant
structures at 4.36 and 4.66~$\gev$~\cite{pppsp}. Recently, both
BaBar and Belle updated their results on $\EE\to \gamma_{\rm ISR}
\pi^+\pi^-J/\psi$, which still show differences in the
$4.008~\gevcs$ mass region~\cite{babary_new,belley_new}; the
latest $e^+e^- \to \gamma_{\rm ISR} \pi^+\pi^-\psi(2S)$
analysis from the BaBar experiment with its full data sample
confirmed the existence of the $Y(4660)$
state~\cite{babar_pppsp_new}. However, in an ISR study of $\EE\to
\eta\jpsi$ by Belle, only the well established $\psift$ and
$\psifto$ charmonium states but no $Y$ states are
observed~\cite{etajpsi}. A better understanding of the structures
observed in these final states would benefit from improved
measurements.

Complementary to the aforementioned neutral states, charged
charmoniumlike structures were observed recently at Belle
and BESIII in the $Y(4260)\to \pi^{\mp}Z(3900)^{\pm}\to \ppjpsi$
decays~\cite{belley_new,zc3900} and at BESIII in $\EE\to
\pi^{\mp}Z_c(4020)^{\pm}\to \pi^\mp (\pi^\pm h_c)$~\cite{zc4020}.
Since these states contain both a $\ccb$ component and
electric charge, they are good candidates for tetraquark or
meson molecular states. These or similar states may exist in
the $\pi^{\pm}\psp$ invariant mass distribution in the $\EE\to
\pp\psp$ process.

To characterize more precisely the properties of the
$Y(4360)$ and $Y(4660)$, to better understand their nature, and to
search for possible charged charmoniumlike states decaying into
$\pi^\pm\psp$, we measure the $\EE\to \pp\psp$ process using
the ISR technique with the full Belle data that was
collected with the Belle detector~\cite{Belle} at the KEKB
asymmetric-energy $e^+e^-$ collider ($3.5~\gev$ $e^+$ and 
$8.0~\gev$ $e^-$)~\cite{KEKB}. The results
here for $Y(4360)$ and $Y(4660)$ supersede our previous measurements
in Ref.~\cite{pppsp}.

In this analysis, $\psp$ is
reconstructed in the $\pp\jpsi$ (hereinafter denoted the ``$\ppjpsi$ mode'')
and the $\MM$ (the ``$\MM$ mode'') final
states and $\jpsi$ is reconstructed in the $\LL~(\ell=e,~\mu)$
final state. Due to the high background from Bhabha
scattering, the $\psp\to \EE$ decay is not used here. The $980~\infb$
data sample used for this analysis was collected at the $\Upsilon(nS)$
($n=1$, 2, 3, 4, or 5) resonances and center-of-mass energies a few
tens of $\mev$ lower than the $\Upsilon(4S)$ or the
$\Upsilon(1S)/\Upsilon(2S)$ peaks.

\section{Detector and Monte Carlo Simulations}

The Belle detector is a large-solid-angle magnetic spectrometer 
that consists of a silicon vertex detector,
a 50-layer central drift chamber, an array of aerogel threshold 
Cherenkov counters, a barrel-like arrangement of time-of-flight scintillation counters,
and an electromagnetic calorimeter comprised of CsI(Tl)
crystals located inside a super-conducting solenoid
coil that provides a 1.5T magnetic field. An iron flux return 
located outside of the coil is instrumented to detect 
$K^0_{\rm L}$ mesons and to identify muons. 
The origin of the coordinate system is defined as the position of the
nominal interaction point. The  $z$ axis is aligned with
the direction opposite the $e^+$ beam and is parallel to the
direction of the magnetic field within the solenoid. The
$x$ axis is horizontal and points towards the outside of the
storage ring and the $y$ axis is vertical upward. The polar
angle and azimuthal angle $\phi$ are measured relative to
the positive $z$ and $x$ axes, respectively.

We use a GEANT-based Monte Carlo (MC) simulation~\cite{geant3} 
to model the response of the detector, identify potential 
backgrounds and determine the acceptance. The MC simulation 
includes run-dependent detector performance variations and 
background conditions.

We use the event generator {\sc phokhara}~\cite{phokhara} to
simulate the process $\EE \to \gamma_{\rm ISR}+Y$. 
In the generator, one or two ISR photons 
may be emitted before forming the resonance $Y$, which then 
decays to $\pp\psp$, with $\psp\to \pp\jpsi\to \pp\LL$ or $\psp\to \MM$.
The masses and widths of $Y(4360)$ and $Y(4660)$ determined in
our previous measurement are used in the simulation~\cite{pppsp}.

\section{Event selection}

For candidate events, we require six (four) well-reconstructed
charged tracks with zero net charge for the $\psp\to \pp\jpsi$
($\psp\to \MM$) mode. Well-reconstructed charged tracks have impact
parameters perpendicular to and along the $e^+$ beam direction with
respect to the interaction point that are less than 0.5~cm and
5.0~cm, respectively. The transverse momentum of each track is
required to be greater than $0.1~\gevc$. For charged tracks,
information from different detector subsystems is combined to form a
likelihood $\mathcal{L}_i$ for particle species
$i$~\cite{pid}. Tracks with $\mathcal{R}_K =
{\mathcal{L}_K}/(\mathcal{L}_K + \mathcal{L}_\pi) < 0.4$ are
identified as pions with an efficiency of 95\%; 6\% of kaons are
misidentified as pions. Similar likelihood ratios are formed for
electron and muon identification~\cite{EID,MUID}. For electrons from
$\jpsi\to \EE$, both tracks are required to have $\mathcal{R}_e >
0.1$. Bremsstrahlung photons detected in the electromagnetic
calorimeter within 0.05~radians of the original lepton
direction are included in the calculation of the $\EE(\gamma)$
invariant mass. For muon candidates in the $\MM$ mode, one
of the tracks is required to have $\mathcal{R}_\mu
> 0.9$ and the other track must have associated hits in the
$K_L$-and-muon detector that agree with the extrapolated trajectory
of a charged track found in the drift chamber. For muons in
$\jpsi\to \MM$, one track must have $\mathcal{R}_\mu>0.9$
but no additional constraints are placed on the other track.

\begin{figure*}[htbp]
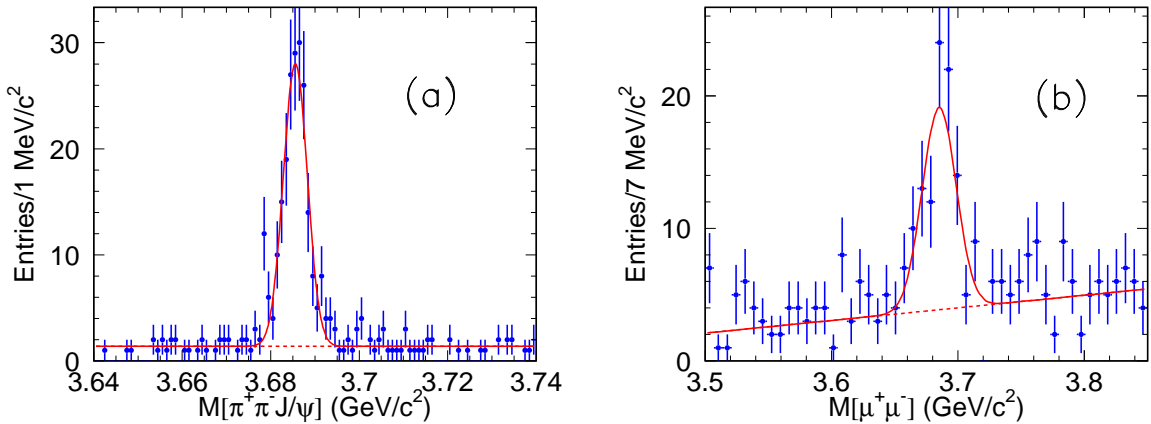

\centerline{
 \psfig{file=mpsp-ppjpsi-draft.epsi,width=8.0cm}
 \psfig{file=mll-mm-ll-draft.epsi,width=8.0cm}}
\caption{Invariant-mass distributions of the candidate $\psp$
signals in (a) the $\ppjpsi$ mode and (b) the $\MM$ mode. Dots
with error bars are data and the curves are the best fits. }
\label{mppll-ppjpsi}
\end{figure*}

For the $\ppjpsi$ mode, there is a clear $\jpsi$ signal in the
lepton-pair invariant-mass distribution. Fitting the mass spectrum
of the lepton pair with a Gaussian function and a linear background,
we obtain an invariant mass of $M_{\LL} = (3099.1\pm
1.7)~\mevcs$ and a resolution ($\sigma_{\LL}$) of $(14.3\pm
1.3)~\mevcs$. The $\jpsi$ signal region is defined as
$m_{\jpsi}-3\sigma_{\LL} < M_{\LL} < m_{\jpsi}+3\sigma_{\LL}$, where
$m_{\jpsi}$ is the nominal world-average $\jpsi$
mass~\cite{PDG}. In the 10\% of events where there
are multiple $\pp$ combinations that satisfy the $\psp$
requirements, we select the one with $|M_{\pp\LL}-M_{\LL}|$,
the difference of the corresponding invariant masses, closest to the
difference of the nominal masses of $\psp$ and $\jpsi$.
Fitting the mass spectrum of the candidate
$\pp\jpsi$-mode events~\cite{mppjpsi} with a
Gaussian function and a linear background, shown in
Fig.~\ref{mppll-ppjpsi}(a), we obtain an invariant mass of
$(3685.4\pm 0.2)~\mevcs$ with a resolution of $(2.7\pm 0.2)~\mevcs$.
The $\psp$ sample is nearly background-free. The $\psp$
signal region is defined as $3.67~\gevcs < M_{\pp\jpsi} <
3.70~\gevcs$, as in our previous measurement~\cite{pppsp}.
The sideband regions are defined as $3.64~\gevcs <
M_{\pp\jpsi} < 3.67~\gevcs$ and $3.70~\gevcs < M_{\ppjpsi} <
3.73~\gevcs$, double the width of the signal region.

For the $\MM$ mode, the invariant-mass distribution of the
$\MM$ pair ($M_{\MM}$) shows a clear $\psp$ signal. From the fit,
shown in Fig.~\ref{mppll-ppjpsi}(b), we obtain an invariant mass of
$(3685.2\pm 2.5)~\mevcs$ with a resolution of $(13.8\pm
2.0)~\mevcs$. The $\psp$ signal region is defined as $3.651~\gevcs < M_{\MM} < 3.721~\gevcs$.
The sideband regions are defined as $3.5215~\gevcs < M_{\MM} <
3.6265~\gevcs$ and $3.7455~\gevcs < M_{\MM} < 3.8505~\gevcs$,
triple the width of the signal region.

For both modes, some $\gamma$ conversions are
misidentified as $\pp$; these events are removed by requiring
$\mathcal{R}_e<0.75$ for the $\pip$ and $\pim$ daughters of the
$\psp$. This background is worse in the $\MM$ mode and so an
invariant mass $M_{\pp}>0.31~\gevcs$ is also
required.

\begin{figure*}[htbp]
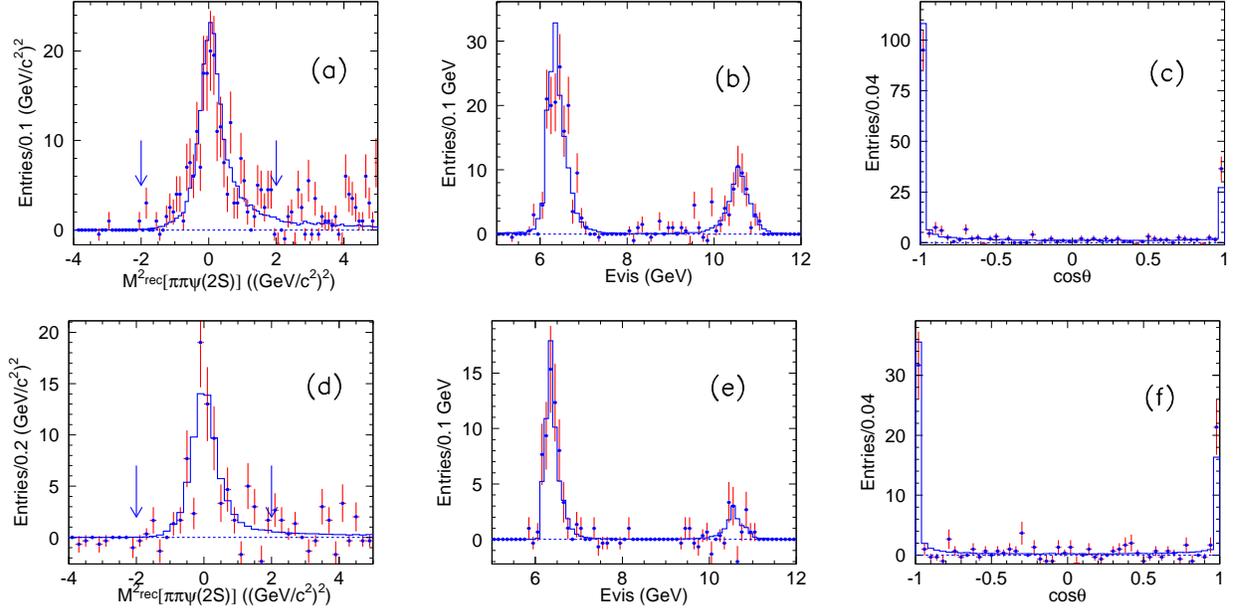

\centerline{
 \psfig{file=mms-ppjpsi-draft.epsi,width=5.5cm}
 \psfig{file=evis-ppjpsi-draft.epsi,width=5.5cm}
 \psfig{file=xcos-ppjpsi-draft.epsi,width=5.5cm}}
\centerline{
 \psfig{file=mms-draft-ll.epsi,width=5.5cm}
 \psfig{file=evis-draft-ll.epsi,width=5.5cm}
 \psfig{file=xcos-draft-ll.epsi,width=5.5cm}
} \caption{The ISR characteristics of the selected events. The first
row is for the $\ppjpsi$ mode and the second for the
$\MM$ mode. Panels (a) and (d) show the square of
the mass recoiling against the $\pp\psp$ system; (b) and (e) show
the visible energy in the detector; (c) and (f) show the
polar-angle distribution of the $\pp\psp$ system in the
$\EE$ center-of-mass frame. Points with error bars
(histograms) represent the data (MC simulation, described in
section~\ref{part1}). The backgrounds, estimated from the
normalized $\psp$ mass sidebands, have been subtracted from the
distributions.} \label{isr-ppjpsi}
\end{figure*}

The detection of the ISR photon ($\gisr$) is optional;
instead, we require $-2.0~(\gevcs)^2 < \MMS < 2.0~(\gevcs)^2$, where
$\MMS$ is the square of the mass recoiling against the $\pp\psp$
system. Good agreement between data and MC simulation
for the visible energy ($E_{\rm vis}$) and
polar-angle distributions of the $\pp\psp$ system in the
$\EE$ center-of-mass frame confirms that the signal events
are produced via ISR. Here, $E_{\rm vis}$ encompasses all
final-state photons and charged particles; energies for the
latter are calculated from track momenta, assuming the tracks to be
pions. The distributions of $\MMS$, $E_{\rm vis}$ and
polar-angle distributions of the $\ppjpsi$ and $\MM$ modes
are shown in Fig.~\ref{isr-ppjpsi}.

\begin{figure*}[htbp]
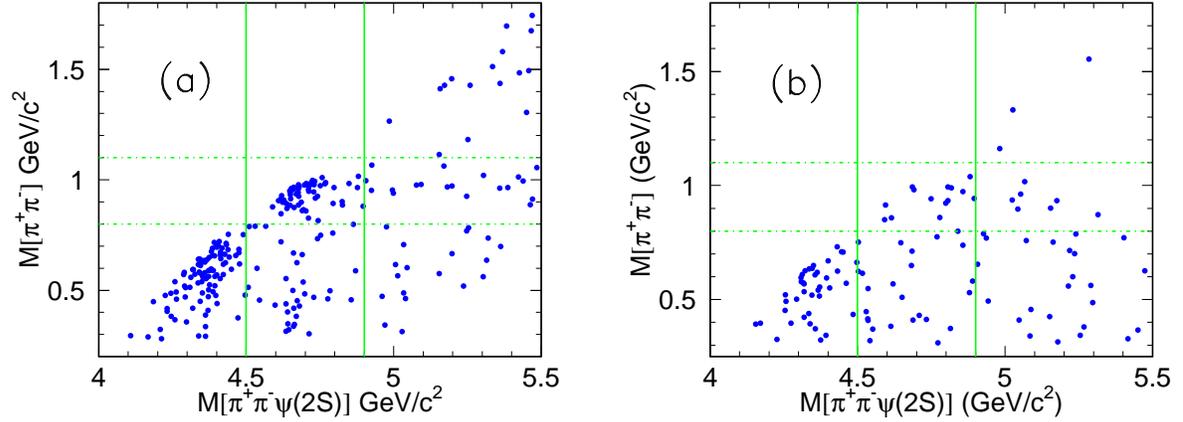

 \centerline{
 \psfig{file=mppmy-ppjpsi-draft.epsi,width=8.0cm}
 \psfig{file=mppmy-ll-draft.epsi,width=8.0cm}}
\caption{Invariant mass of the $\pp$ recoiling against the $\psp$
 versus the invariant mass of the $\pp\psp$
 in the $\ppjpsi$ mode (a) and $\MM$ mode
(b). The horizontal dashed lines show the belt of $\fz$,
while the vertical solid lines demarcate the regions
with the $Y(4360)$ and $Y(4660)$ states and
the higher-mass combinations. }\label{mppmy}
\end{figure*}

\begin{figure}[htbp]
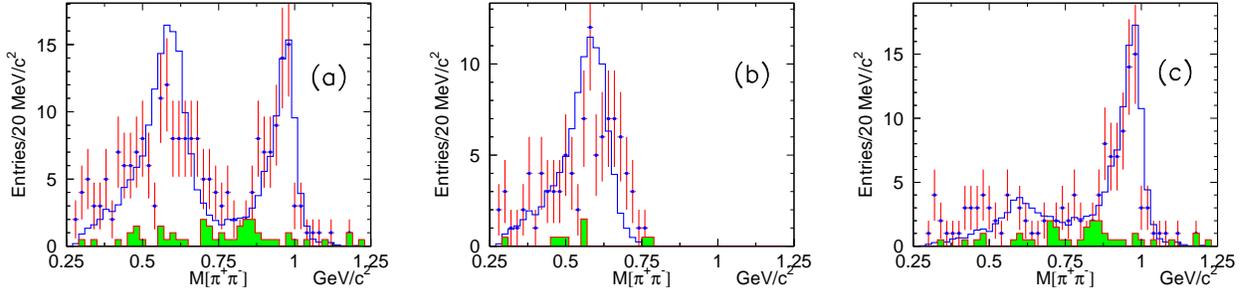

\centerline{
 \psfig{file=mpp-tot-ppjpsi-draft.epsi,width=5.5cm}
 \psfig{file=mpp-y43-ppjpsi-draft.epsi,width=5.5cm}
 \psfig{file=mpp-y46-ppjpsi-draft.epsi,width=5.5cm}}
\caption{Invariant mass distributions of $\pp$ from $Y(4360)$ and
$Y(4660)$ decays in the $\ppjpsi$ mode.
Points with error bars (open histograms) represent the data
(MC simulation, described in section I); the shaded histograms
represent the background estimated from the scaled sidebands.  Panel 
 (a) is  for the events in the
region $4.0~\gevcs<M_{\pp\psp}<5.5~\gevcs$, (b) for events
in the $Y(4360)$ region (4.0 to $4.5~\gevcs$), and (c)
for the events in $Y(4660)$ region (4.5 to $5.5~\gevcs$). }
\label{m2piy-ppjpsi}
\end{figure}

After all the above selections, there are 245 $\pp\psp$ candidate
events with 28 background events in the $\pp\jpsi$ mode, and 118 candidate events with 56
background events in the $\MM$ mode; the background yields are
estimated from the corresponding sidebands. Figure~\ref{mppmy} shows the scatter plots of the
invariant mass $M_{\pp}$ of the $\pp$ pair recoiling against the $\psp$
 versus the invariant mass $M_{\pp\psp}$ of the $\pp\psp$ combination~\cite{mpppsp}.
The corresponding distribution of the candidate events in
the $\MM$ mode is similar but with lower statistics. There are two
clusters of events corresponding to the $Y(4360)$ and $Y(4660)$. The
$M_{\pp}$ distributions tend to cluster around the masses of
$f_0(500)$ and $\fz$.  Figure~\ref{m2piy-ppjpsi} shows the
projection onto the $M_{\pp}$ axis in the cleaner
$\pp\jpsi$ mode, compared with MC simulation that assumes an
incoherent sum of the $f_0(500)$ and $\fz$. 
Additionally, the angular distributions of $\pi^{\pm}$ and 
$\pp$ pair are compared. Figure~\ref{cos-pi} shows the angular 
distributions of $\pi$ in the $\pp$ system, and Fig.~\ref{cos-pp} 
shows the $\pp$ in $\pp\psp$ system. The data are from the clean $\ppjpsi$ 
mode, and the MC simulations are generated assuming $S-$wave between $\pip$ 
and $\pim$ in $\pp$ system and $S-$wave between $\pp$ and $\psp$ in 
$\pp\psp$ system.

\begin{figure}[htbp]
\centerline{
 \psfig{file=cosp-pip-tot-prd.epsi,width=5.5cm}
 \psfig{file=cosp-pip-y4360-prd.epsi,width=5.5cm}
 \psfig{file=cosp-pip-y4660-prd.epsi,width=5.5cm}}
\caption{Angular distributions of $\pi$ in the $\pp$ system
in $\ppjpsi$ mode. Panel
 (a) is  for the events in the
region $4.0~\gevcs<M_{\pp\psp}<5.5~\gevcs$, (b) for events
in the $Y(4360)$ region (4.0 to $4.5~\gevcs$), and (c)
for the events in $Y(4660)$ region (4.5 to $4.9~\gevcs$). The
dots with error bars are data and the histograms are from MC
simulation.} \label{cos-pi}
\end{figure}

\begin{figure}[htbp]
\centerline{
 \psfig{file=cospp-tot-prd.epsi,width=5.5cm}
 \psfig{file=cospp-y4360-prd.epsi,width=5.5cm}
 \psfig{file=cospp-y4660-prd.epsi,width=5.5cm}}
\caption{Angular distributions of the $\pp$ in $\pp\psp$ system
in the $\ppjpsi$ mode. Panel
 (a) is  for the events in the
region $4.0~\gevcs<M_{\pp\psp}<5.5~\gevcs$, (b) for events
in the $Y(4360)$ region (4.0 to $4.5~\gevcs$), and (c)
for the events in $Y(4660)$ region (4.5 to $4.9~\gevcs$). The
dots with error bars are data, while the histograms are MC
simulation.} \label{cos-pp}
\end{figure}

\section{\boldmath Fit to $M_{\pp\psp}$ and measurement of cross sections}

\begin{figure*}[htbp]
 \psfig{file=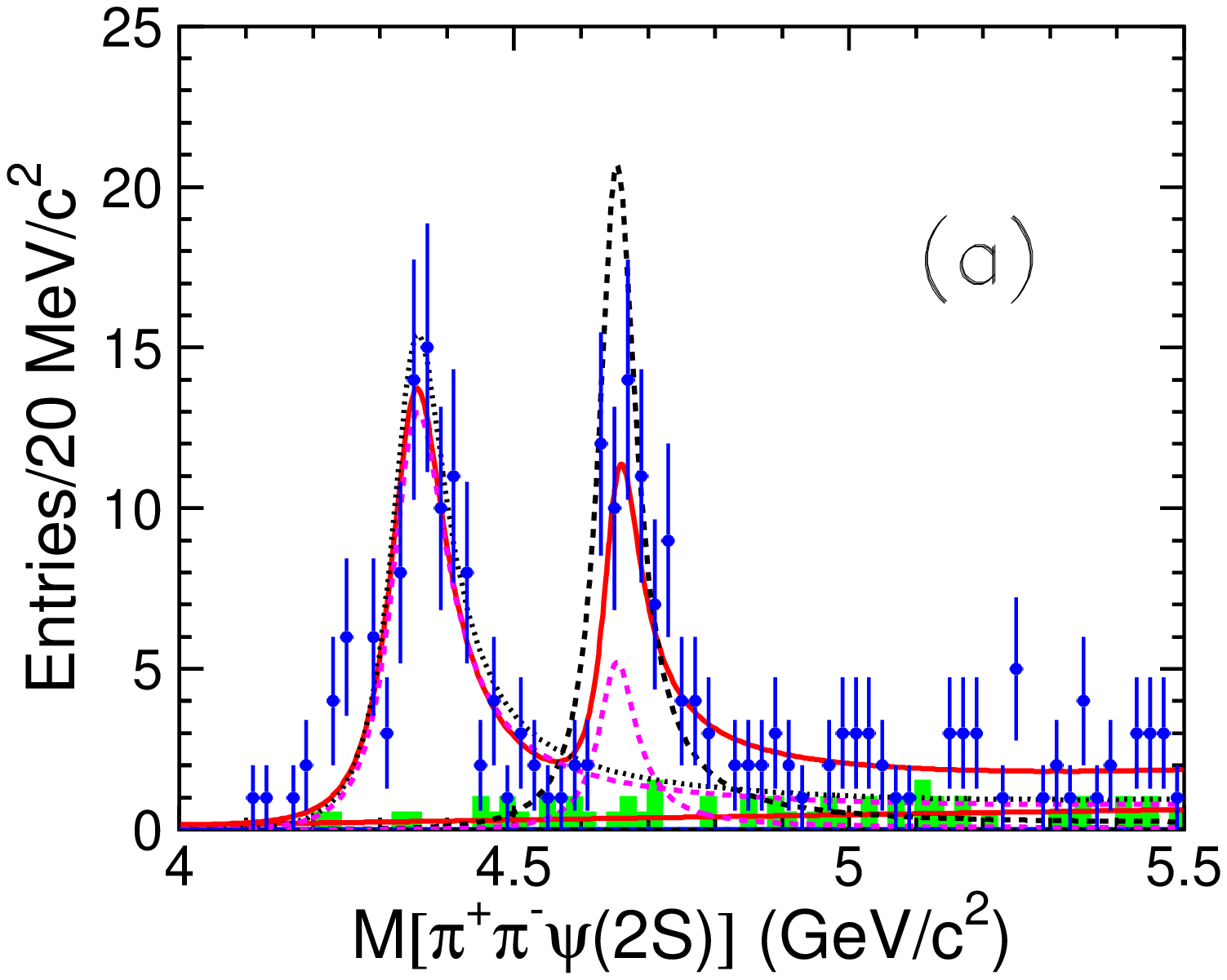, height=5.cm, angle=-90}
 \psfig{file=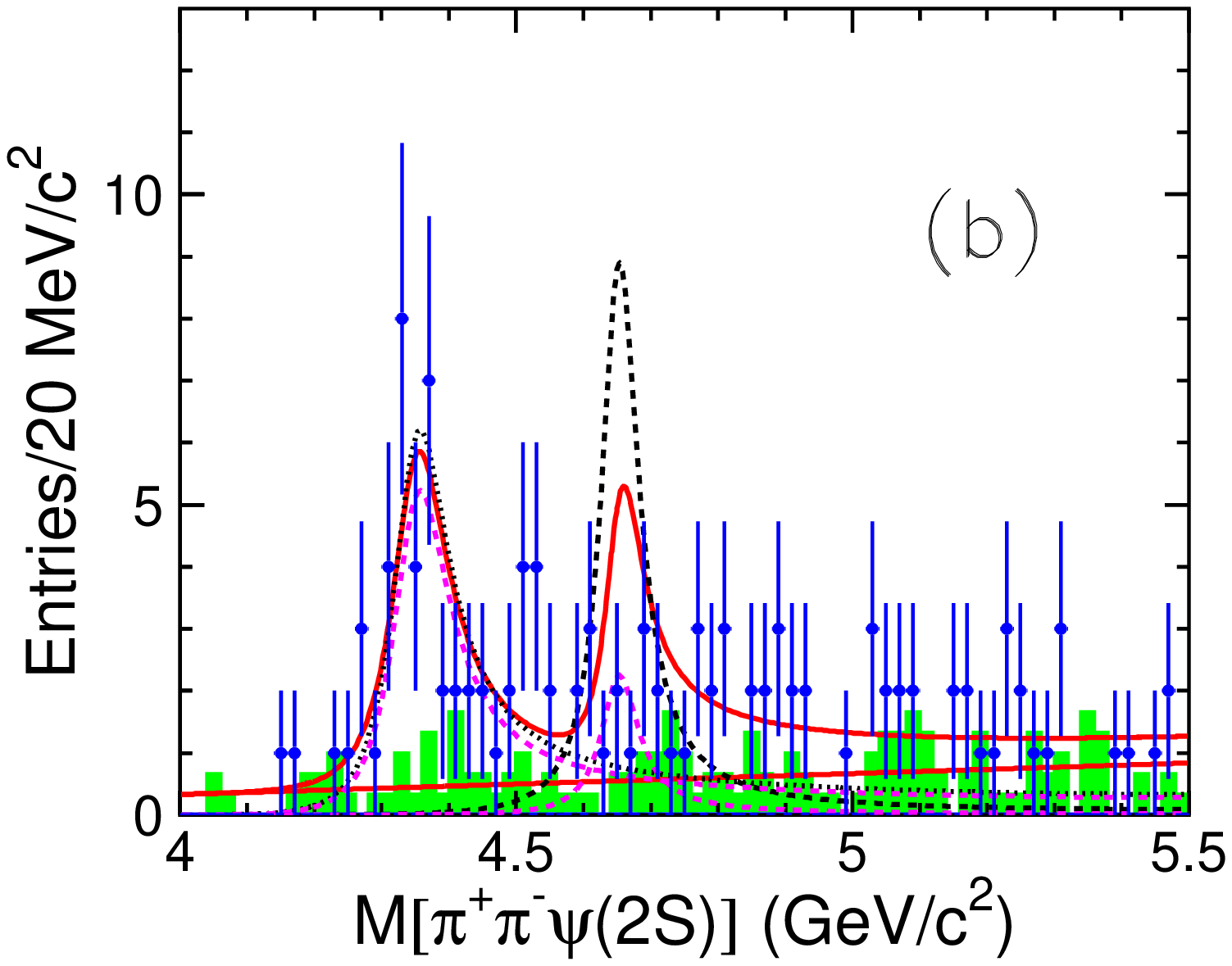, height=5.cm, angle=-90}
 \psfig{file=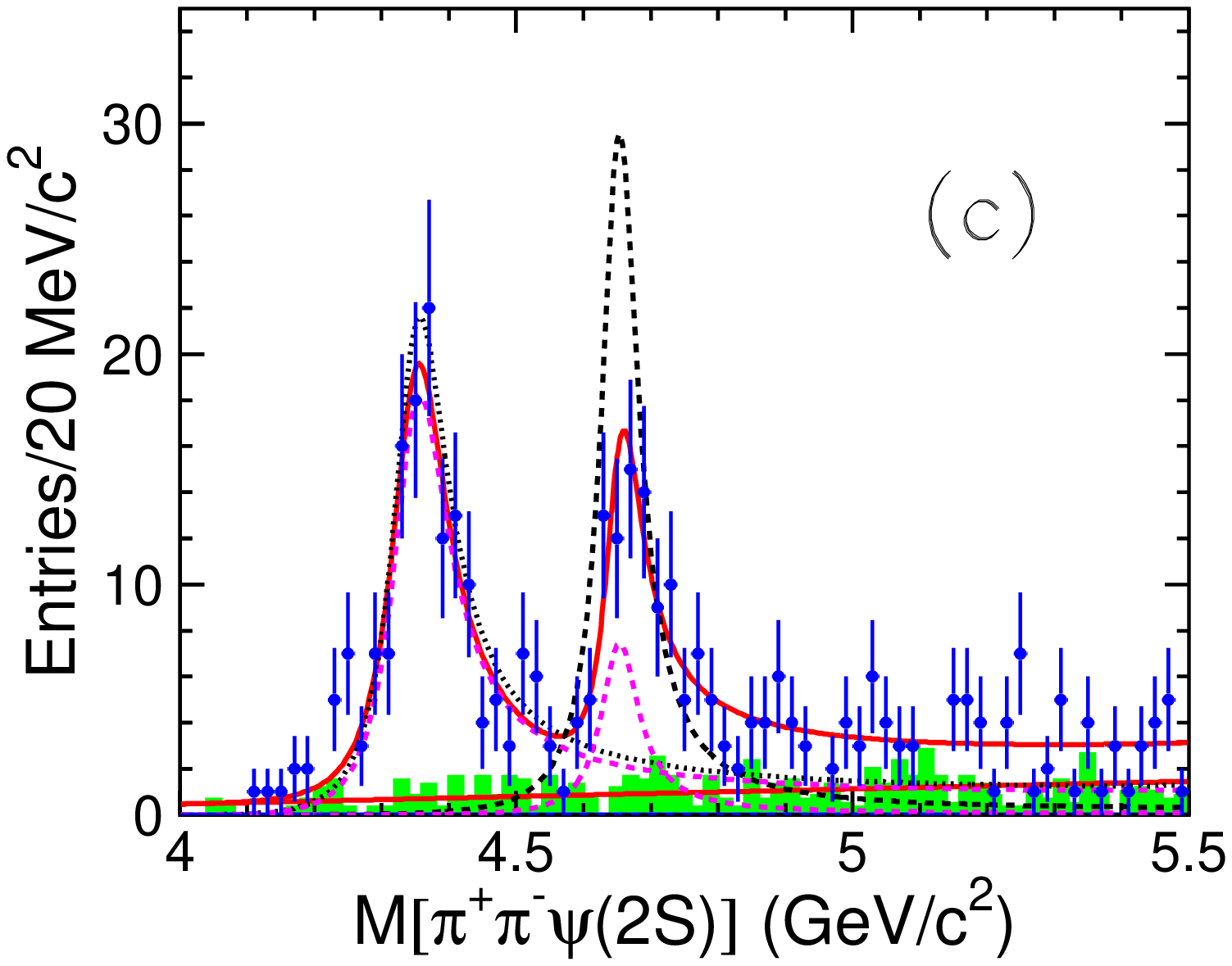, height=5.cm, angle=-90}
\caption{The $\pp\psp$ invariant-mass distributions and the
simultaneous fit results described in the text. From left to
right: (a) the $\ppjpsi$ mode, (b) the $\MM$ mode,
and (c) the sum. The points with error bars show the data while the
shaded histograms are the scaled sideband
backgrounds. The solid red curves show the best fits; the dashed curves,
which are from the two fit solutions, show the contributions from the
two Breit-Wigner components (described in the text). The
interference between the two resonances is not shown. }
\label{fit}
\end{figure*}

Figure~\ref{fit} shows the $M_{\pp\psp}$ distributions in the
$\ppjpsi$ and $\MM$ modes; the structures in these two modes agree
with each other within statistics. To extract the resonant
parameters of the two $Y$ states, an unbinned maximum-likelihood fit
is performed to the mass spectra $M_{\pp\psp}\in [4.0,5.5]~\gevcs$
simultaneously for the $\ppjpsi$ and $\MM$ modes,
assuming that only two resonances and an incoherent featureless
background contribute. The $\psp$ mass-sidebands are included in
the fit to estimate the backgrounds in the signal region;
here, the fit assumes only the background component. The fit to the
events in the signal region includes two coherent $P$-wave
Breit-Wigner  functions, $f_1$ for the $Y(4360)$ and
$f_2$ for the $Y(4660)$. 

The amplitude of the Breit-Wigner function $f_j$ ($j=1,2$)
is defined as
\begin{equation}
 f_j(M_{\pp\psp}) =\frac{M_j}{M_{\pp\psp}}
\frac{\sqrt{12\pi\mathcal{B}_j(\pp\psp)\Gamma_j^{\EE}\Gamma_j}}
{M^2_{\pp\psp}-M_j^2+iM_j\Gamma_j}\cdot
\sqrt{\frac{\Phi(M_{\pp\psp})}{\Phi(M_j)}}, \label{bw}
\end{equation}
where $\Gamma_j^{\EE}$ is the partial width to $e^+e^-$, $\Gamma_i$
the total width that is assumed to be a constant, and
$\mathcal{B}_j(\pp\psp)$ the branching fraction of the resonance's
decay to $\pp\psp$. $\Phi(m)$ is the three-body phase-space factor
for a resonance of mass $m$ that decays to $\pp\psp$.
 In the fit, $M_j$, $\Gamma_j$, and the product
$\mathcal{B}_j(\pp\psp)\Gamma_j^{\EE}$ are free parameters.

\begin{figure}[htbp]
 \psfig{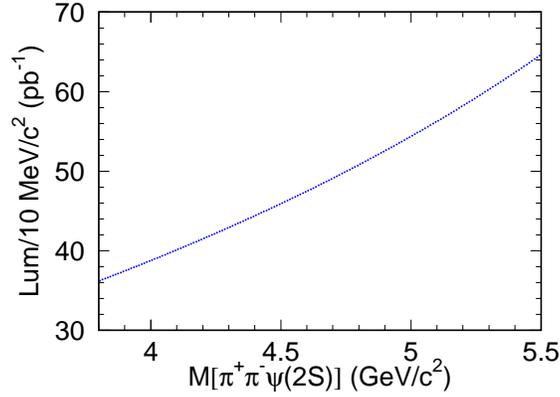}
\caption{The effective luminosity of ISR production with the full data sample.}
\label{lum-eff}
\end{figure}

\begin{figure}[htbp]
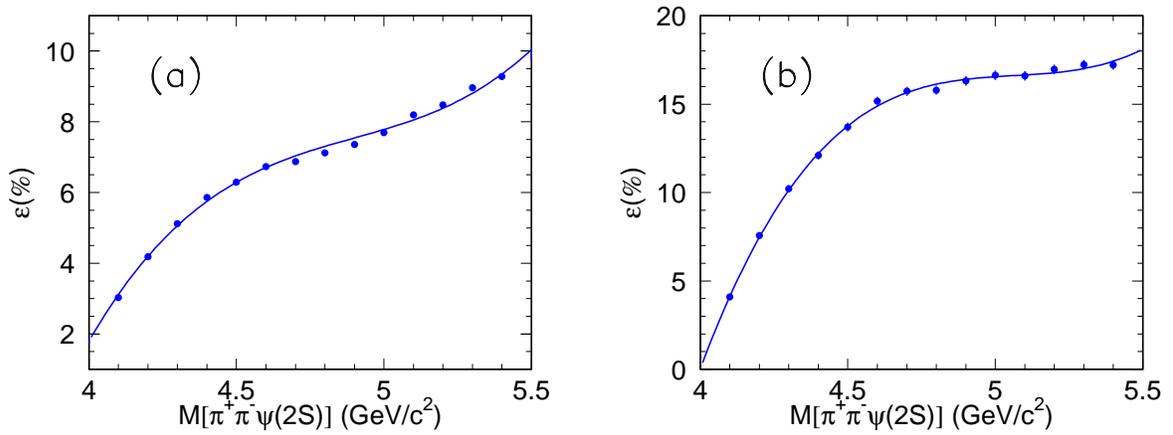

 \psfig{file=./eff-ppjpsi-bn.epsi,width=8cm}
 \psfig{file=./eff-ll-bn.epsi,width=8cm}
\caption{The efficiency curves and the fit to third order polynomials. 
Plot (a) is $\ppjpsi$ mode, and (b) is $\MM$ mode.} \label{eff-curve}
\end{figure}

The signal amplitude is $A = f_1+f_2\cdot e^{i\phi}$, where $\phi$
is the relative phase between the two resonances,
and the $M_{\pp\psp}$ distribution of
signal events is then $\mathcal{L}_{eff}\cdot\eff\cdot |A|^2$. 
Here the $\mathcal{L}_{eff}$ is the effective
luminosity~\cite{kuraev} and $\eff$ is the $M_{\pp\psp}$-dependent efficiency.
The effective luminosity of ISR is calculated according to
theoretical formulae~\cite{kuraev} and the integrated luminosity of 
Belle data, which is shown in Fig.~\ref{lum-eff}. To determine the 
efficiency in the range of $4.0~<M_{\pp\psp}<5.5~\gevcs$, MC samples 
with different $M_{\pp\psp}$ are generated and simulated. The efficiency 
curves are shown in Fig.~\ref{eff-curve}.
The mass resolution, which is determined from MC simulation to range
from 2 to $5~\mevcs$ over the fit region, is small compared
with the widths of the observed structures and so is
neglected. The fit results are shown in Fig.~\ref{fit} and
Tables~\ref{two_sol}~and~\ref{corr}. There are two solutions with
equally good fit quality; the $\chi^2/ndf$ is $18.7/21$, where $ndf$
is the number of degrees of freedom.

\begin{table}
\caption{Results of the fits to the $\pp\psp$ invariant-mass
spectra. The first error statistical and the second
systematic. $M$, $\Gamma$, and $\BR\cdot \Gamma^{\EE}$ are the
mass (in $\mevcs$), total width (in $\mev$), and the product of the
branching fraction to $\pp\psp$ and the $\EE$ partial width (in
eV), respectively; $\phi$ is the relative phase between the two
resonances (in degrees).}\label{two_sol}
\begin{center}
\begin{tabular}{c |c c }
\hline\hline
     Parameters     & ~~~Solution~I~~~ & ~~~Solution~II~~~  \\\hline
 $M_{Y(4360)}$        & \multicolumn{2}{c}{$4347\pm6\pm3$} \\
 $\Gamma_{Y(4360)}$   & \multicolumn{2}{c}{$103\pm9\pm5$}  \\
 $\BR[Y(4360)\to\pp\psp]\cdot\Gamma_{Y(4360)}^{\EE}$
                    & ~~$9.2\pm0.6\pm0.6$~~ & ~~$10.9\pm0.6\pm0.7$~~   \\
 $M_{Y(4660)}$       & \multicolumn{2}{c}{$4652\pm10\pm11$} \\
 $\Gamma_{Y(4660)}$  & \multicolumn{2}{c}{$68\pm11\pm5$}   \\
 $\BR[Y(4660)\to\pp\psp]\cdot\Gamma_{Y(4660)}^{\EE}$
                    & ~~$2.0\pm0.3\pm0.2$~~ & ~~$8.1\pm1.1\pm1.0$~~  \\
 $\phi$             & ~~$32\pm18\pm20$~~ & ~~$272\pm8\pm7$~~  \\
 \hline\hline
\end{tabular}
\end{center}
\end{table}

\begin{table}
\caption{The correlations between the fit parameters shown in
Table~\ref{two_sol} (with the units given there).
The numbers in parentheses are for the
second solution.}\label{corr}
\begin{center}
\begin{tabular}{c |c c c c c c c}
\hline\hline
          & $\Gamma_{Y(4360)}$ & $\BR\cdot\Gamma_{Y(4360)}^{\EE}$
          & $M_{Y(4660)}$ & $\Gamma_{Y(4660)}$ & $\BR\cdot\Gamma_{Y(4660)}^{\EE}$
          & $\phi$   \\\hline
 $M_{Y(4360)}$         & -0.34~(-0.34) & 0.04~(0.04) & -0.29~(-0.29) & 0.05~(0.05) & 0.30~(-0.13) & -0.37~(0.36) \\
 $\Gamma_{Y(4360)}$    & 1.00 & 0.12~(0.12) & -0.08~(-0.08) & -0.28~(-0.28) & -0.45~(-0.11) & -0.08~(-0.10) \\
 $\BR\cdot\Gamma_{Y(4360)}^{\EE}$
                       & -- & 1.00 & -0.37~(-0.22) & -0.32~(0.01) & -0.28~(0.03) & -0.40~(0.06)  \\
 $M_{Y(4660)}$         & -- & -- & 1.00 & 0.21~(0.21) & -0.06~(0.54) & 0.86~(-0.76) \\
 $\Gamma_{Y(4660)}$    & -- & -- & -- & 1.00 & 0.14~(0.74) & 0.25~(-0.44)  \\
 $\BR\cdot\Gamma_{Y(4660)}^{\EE}$
                       & -- & -- & -- & -- & 1.00 & -0.17~(-0.72) \\
 \hline\hline
\end{tabular}
\end{center}
\end{table}

\begin{figure*}[htbp]
 \psfig{file=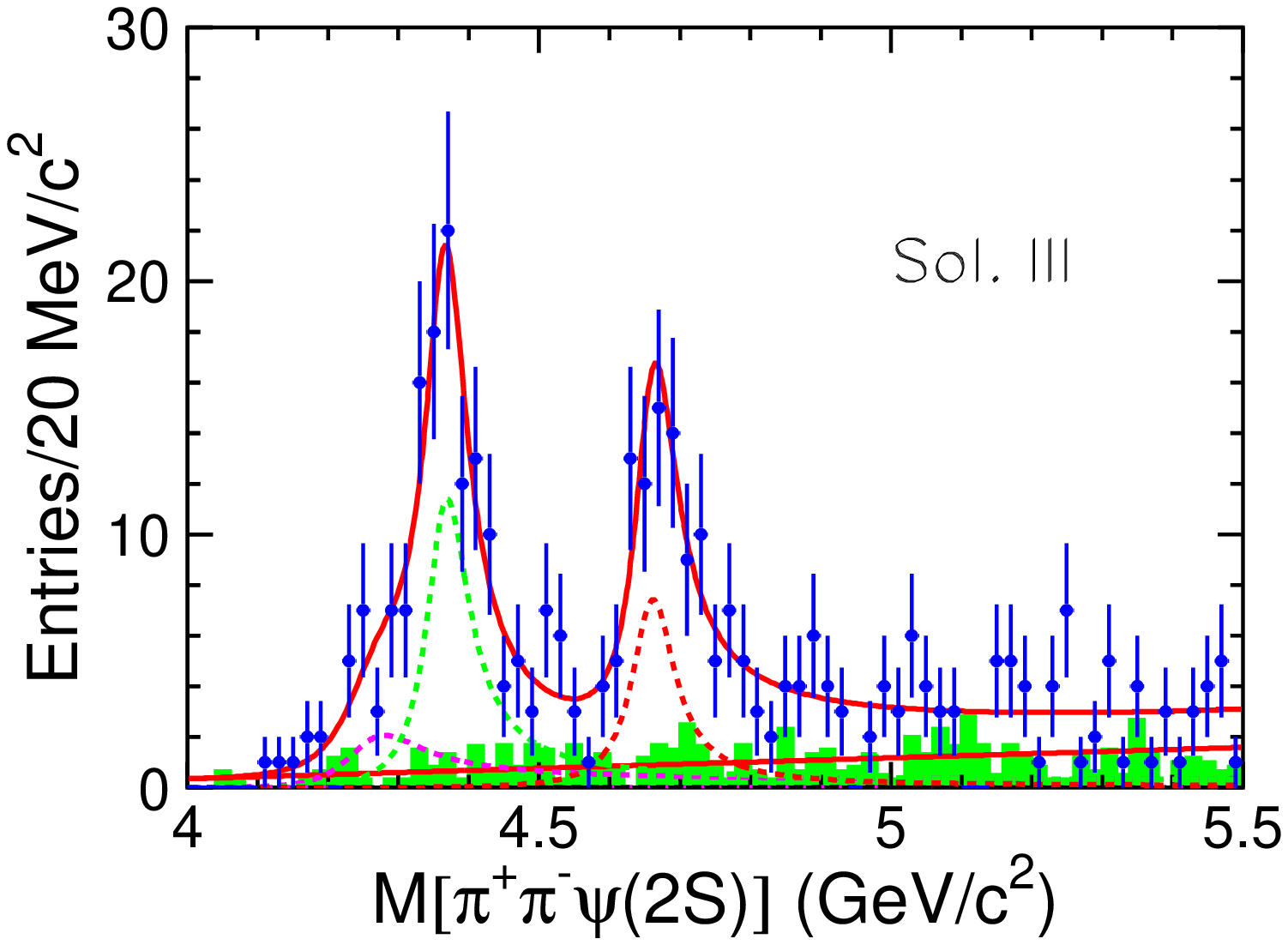, height=8.cm, angle=-90}
 \psfig{file=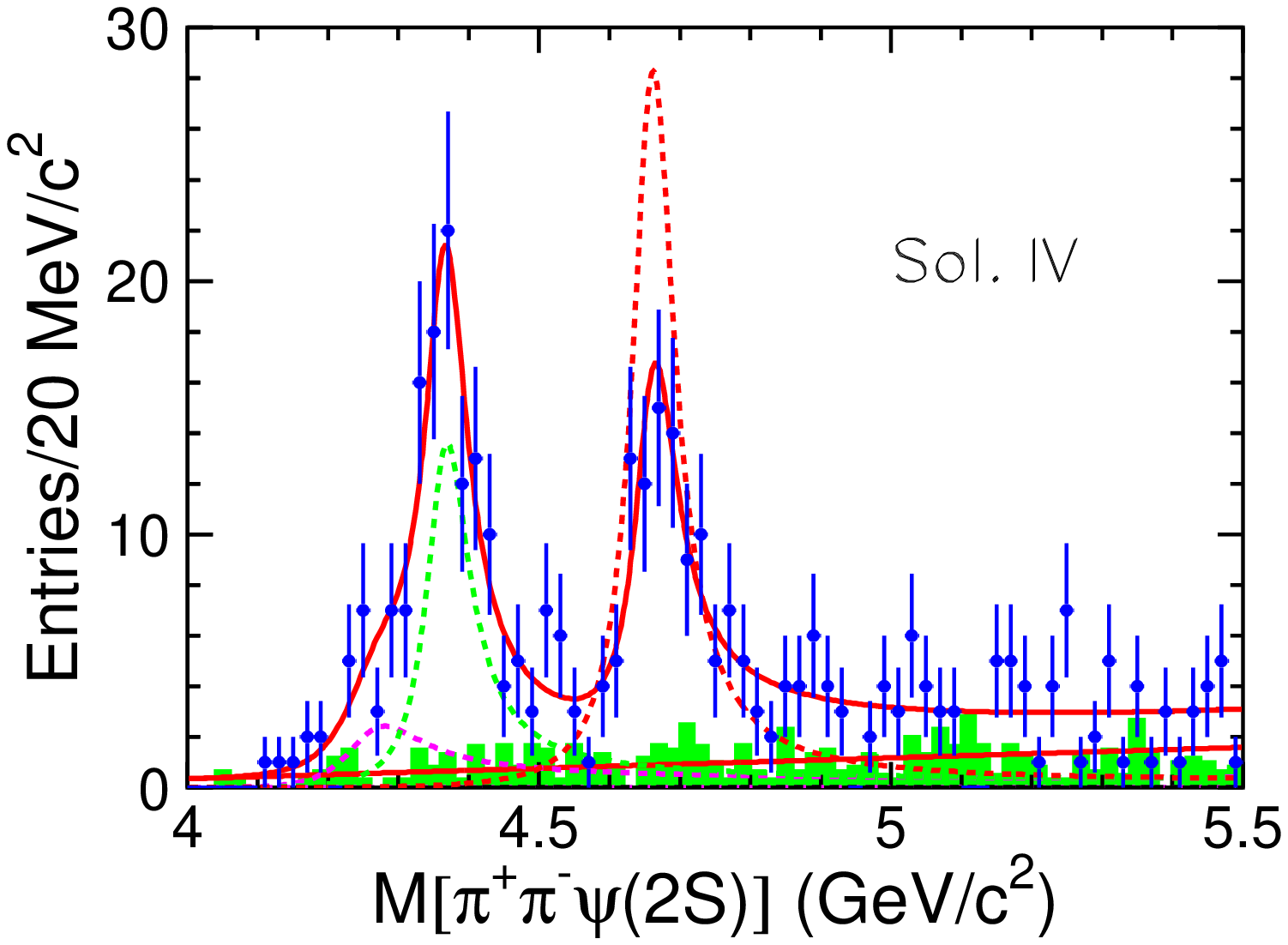, height=8.cm, angle=-90}\\
 \psfig{file=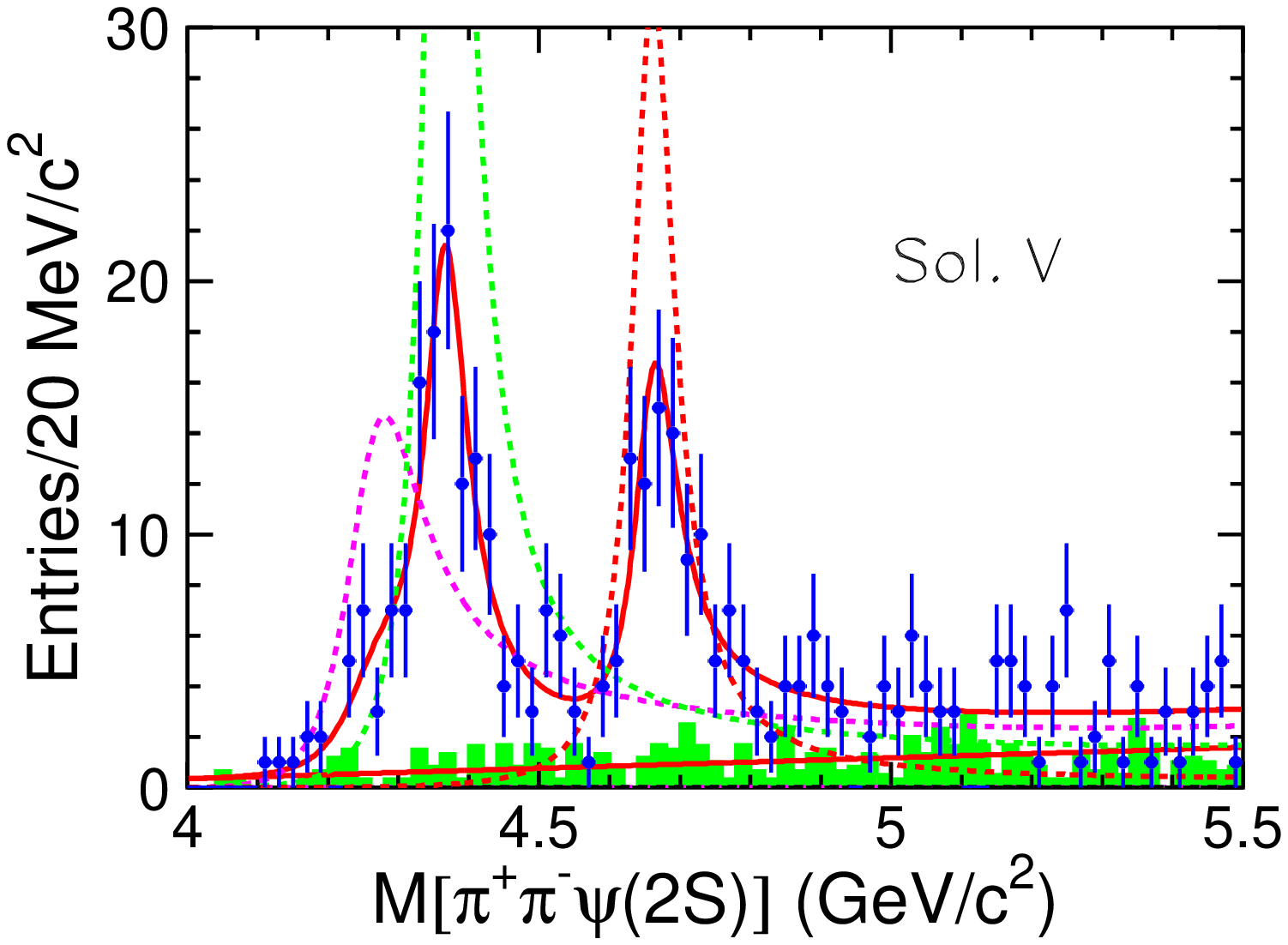, height=8.cm, angle=-90}
 \psfig{file=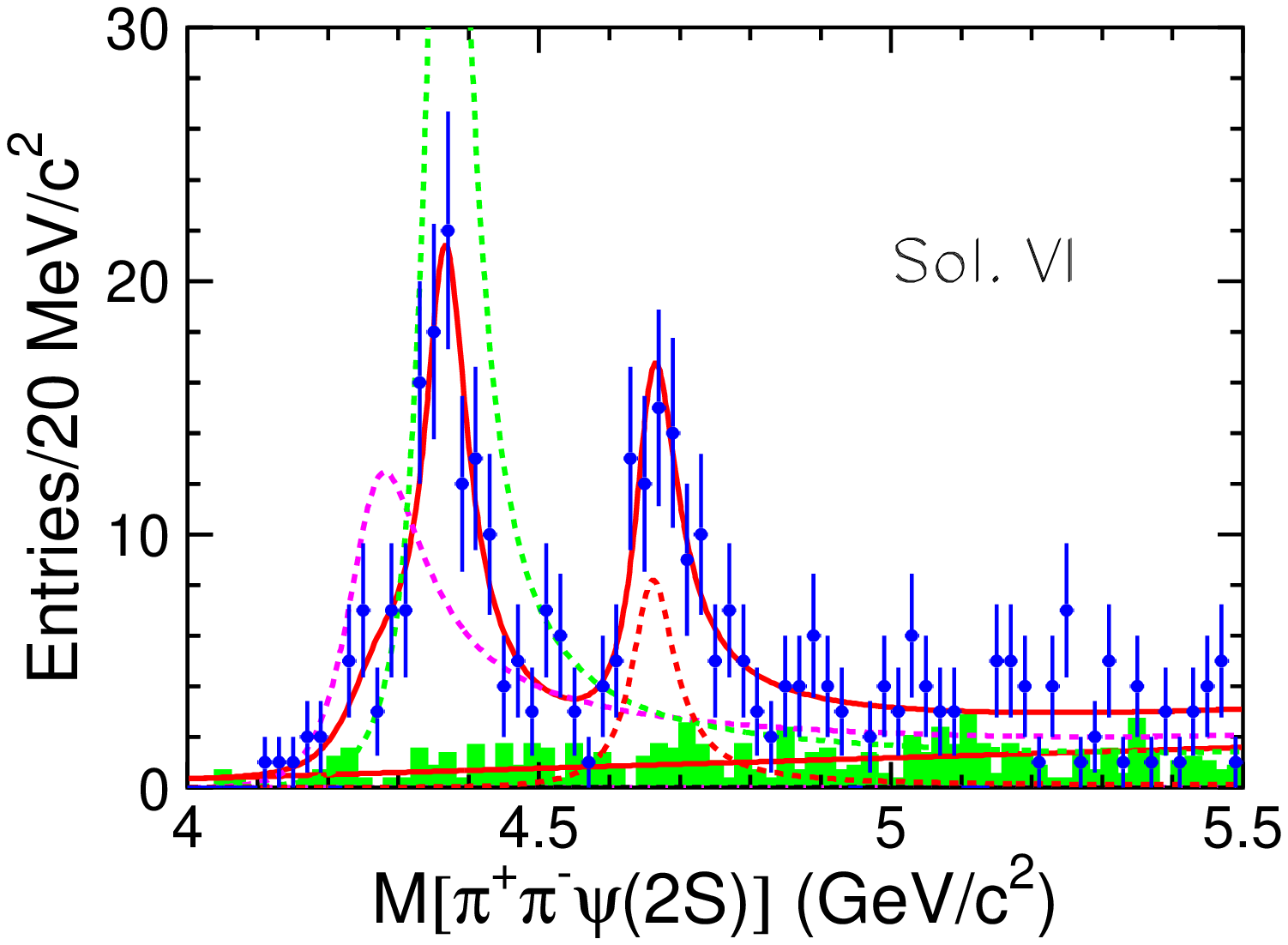, height=8.cm, angle=-90}
\caption{The four solutions from the fit to the $\pp\psp$
invariant mass spectra with the $Y(4260)$ included. The curves
show the best fit and the dashed curves show the contributions
from the two Breit-Wigner components. } \label{fit-3r}
\end{figure*}

Since there are a number of events in the vicinity of the
$Y(4260)$ mass, an alternative fit with a coherent
sum of $Y(4260)$, $Y(4360)$, and $Y(4660)$ amplitudes is performed.
In this fit, the mass and total width of the $Y(4260)$ state
are fixed to their latest measured values~\cite{belley_new}.
There are four solutions with equally good fit quality: $\chi^2/ndf
= 14.8/19$. The signal significance of the $Y(4260)$ is estimated to
be $2.4\sigma$ by comparing the likelihood difference when
the $Y(4260)$ is included in or excluded from the fit. The
fit results are shown in Fig.~\ref{fit-3r} and
Table~\ref{three_res}. Since this significance is marginal,
the solutions without $Y(4260)$ are taken as the nominal
results.

\begin{table*}
\caption{Results of the alternative fits to the $\pp\psp$ invariant-mass
spectra using three resonances: $Y(4260)$, $Y(4360)$, and
$Y(4660)$. The parameters are the same as in Table~\ref{two_sol},
except that, here, $\phi_1$ is the relative phase between the $Y(4360)$
and $Y(4260)$ (in degrees) and $\phi_2$ is the relative phase
between the $Y(4360)$ and $Y(4660)$ (in degrees).
}\label{three_res}
\begin{center}
\begin{tabular}{c |c c c c }
\hline\hline
     Parameters     & ~~Solution~III~~~ & ~~~Solution~IV~~ & ~~Solution~V~~ & ~~~Solution~VI~~ \\\hline
 $M_{Y(4260)}$        & \multicolumn{4}{c}{4259~(fixed)} \\
 $\Gamma_{Y(4260)}$   & \multicolumn{4}{c}{134~(fixed)}  \\
 $\BR[Y(4260)\to\pp\psp]\cdot\Gamma_{Y(4260)}^{\EE}$
                    & $1.5\pm0.6\pm0.4$ & $1.7\pm0.7\pm0.5$ & $10.4\pm 1.3\pm0.8$ & $8.9\pm 1.2\pm0.8$  \\
 $M_{Y(4360)}$        & \multicolumn{4}{c}{$4365\pm7\pm4$} \\
 $\Gamma_{Y(4360)}$   & \multicolumn{4}{c}{$74\pm14\pm4$}  \\
 $\BR[Y(4360)\to\pp\psp]\cdot\Gamma_{Y(4360)}^{\EE}$
                    & $4.1\pm1.0\pm0.6$ & $4.9\pm1.3\pm0.6$ & $21.1\pm 3.5\pm1.4$ & $17.7\pm 2.6\pm1.5$ \\
 $M_{Y(4660)}$       & \multicolumn{4}{c}{$4660\pm9\pm12$} \\
 $\Gamma_{Y(4660)}$  & \multicolumn{4}{c}{$74\pm12\pm4$}   \\
 $\BR[Y(4660)\to\pp\psp]\cdot\Gamma_{Y(4660)}^{\EE}$
                    & $2.2\pm0.4\pm0.2$ & $8.4\pm0.9\pm0.9$ & $9.3\pm 1.2\pm1.0$ & $2.4\pm 0.5\pm0.3$\\
 $\phi_1$             & $304\pm24\pm21$ & $294\pm25\pm23$ & $130 \pm 4\pm2$ & $141\pm 5\pm4$\\
 $\phi_2$             & $26\pm19\pm10$ & $238\pm14\pm21$ & $329\pm 8\pm5$ & $117\pm 23\pm25$\\
 \hline\hline
\end{tabular}
\end{center}
\end{table*}

To compare with our previous measurement~\cite{pppsp}, the
fit to the $\pp\jpsi$ mode alone is performed. The
differences can be explained by the strong correlation between the
parameters (see Table~\ref{corr}). For this mode alone, we
also compare the alternative fit including the $Y(4260)$ with
the nominal fit and consistent results with a $2.8\sigma$
statistical significance for the $Y(4260)$ signal.
The results are discussed further in Appendix~\ref{App:Appendix0}.

\begin{figure}[htbp]
 \psfig{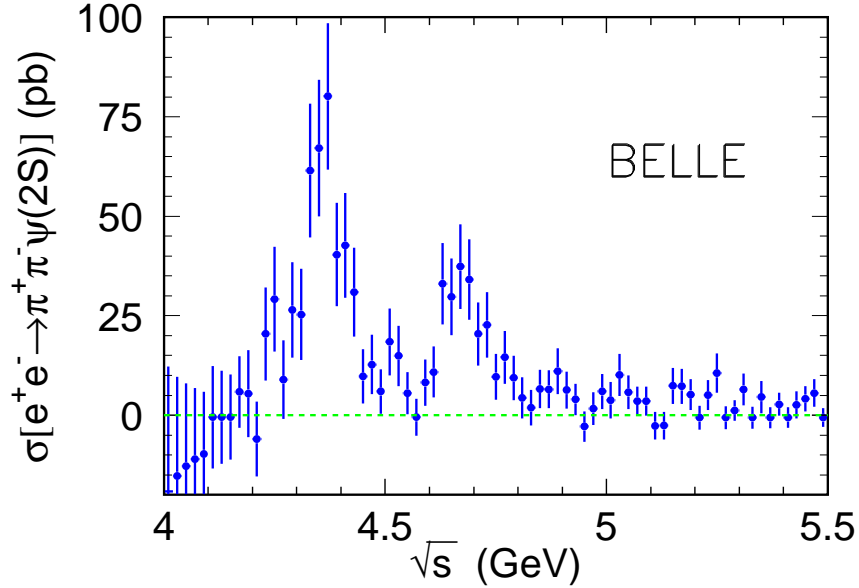}
\caption{The measured $\EE\to \pp\psp$ cross section for
$\sqrt{s}=4.0$ to $5.5~\gev$. The errors are the sum in quardrature of the summed
statistical errors of the numbers of signal and background events
and the systematic errors. } \label{xs_full}
\end{figure}

The invariant mass distributions of the two modes are combined together. 
The cross section for $\EE\to \pp\psp$ in each $\pp\psp$ mass bin
is calculated according to
 \begin{equation*}\label{xscal}
  \sigma_{i} = \frac{n^{\rm obs}_{i} - n^{\rm bkg}_{i}}
    { \lum_i\sum_{j=1}^{2}\eff_{ij}\BR_j},
 \end{equation*}
where $j$ identifies the decay mode of $\psp$
($j=1 $ for the $\pp\jpsi$ mode and $j = 2$ for the
$\MM$ mode) and $i$ indicates the mass bin; $n^{\rm
obs}_{i}$, $n^{\rm bkg}_{i}$, $\eff_{i}$, $\lum_i$,
and $\BR$ are the number of events observed in data, the
number of background events estimated from the fit to the events in
the sidebands and scaled to the signal region, the
detection efficiency of the $j^{\textrm{th}}$ mode, the
effective luminosity in the $i^{\textrm{th}}$ $\pp\psp$ mass
bin, and the branching fractions of the  $j^{\textrm{th}}$
mode~\cite{PDG}, respectively. The resulting cross sections in 
the full solid angle are shown in Fig.~\ref{xs_full} and
Appendix~\ref{App:AppendixA}, where the error bars include
statistical uncertainties in the signal and the subtracted
background and all the systematic errors. The systematic
error for the cross-section measurement is 4.8\% and is the
same for all data points.

\section{Systematic errors}\label{systematic}

The systematic uncertainties in the cross-section measurements are
discussed below.

The particle identification uncertainty is 3.3\% for the
$\ppjpsi$ mode and 1.4\% for the $\MM$ mode. The uncertainty
in the tracking efficiency is 0.35\% per track and is additive. The
efficiency differences between data and MC due to the
corresponding resolutions in the $\jpsi$ mass, $\psp$ mass,
and $\MMS$ requirements are measured with the control sample $\EE\to
\psp\to \ppjpsi$~\cite{belley_new}. The MC efficiency is found to be
higher than in data by $(4.3\pm 0.7)$\% for the $\ppjpsi$ mode and
$(4.4\pm 0.3)$\% for the $\MM$ mode. A correction factor of 1.043
(1.044) is applied to the $\ppjpsi$ ($\MM$) mode, leaving
0.7\% (0.3\%) as the residual systematic error.

The luminosity uncertainty of 1.4\% is due mainly to
the uncertainty from the Bhabha generator. The trigger efficiency
for the events surviving the selection criteria is $(98.7\pm
0.1({\rm stat.}))\%$ for the $\ppjpsi$ mode and $(91.4\pm 0.6({\rm
stat.}))\%$ for the $\MM$ mode, based on the trigger
simulation. A value of 1.0\% is taken as a conservative
estimate of the systematic error for the $\ppjpsi$ mode;
$1.5\%$ is used for the $\MM$ mode.

Uncertainties in the simulation of the ISR process with {\sc phokhara} contributes
less than 1.0\%, and the largest uncertainty in the MC generation
of signal events is from the simulation of the $M_{\pp}$ from $Y$
decays. We generate another MC sample with $m_{f_0(500)} =
0.7~\gevcs$ and $\Gamma_{f_0(500)} = 0.2~\gev$ in order to check
the efficiency variation. The efficiency changes by 2.0\% at
$4.4~\gevcs$ and 3.8\% at $4.7~\gevcs$; half of the larger
efficiency difference, 1.9\%, is taken as the systematic error.
The possible existence of the $Z_c$ structure in $\pi^{\pm}\psp$ 
system doesnot affect the efficiency significantly and is thus neglected.

The uncertainties in the intermediate decay branching fractions
taken from Ref.~\cite{PDG} contribute systematic errors of 1.0\%
for the $\ppjpsi$ mode and 10.4\% for the $\MM$ mode. The
statistical error in the MC determination of the efficiency is
less than 0.1\%.

Assuming all the sources are independent and adding them in
quadrature, we obtain total systematic errors in the cross section
measurement of 5.0\% for the $\ppjpsi$ mode and 11.0\% for the
$\MM$ mode. The combined systematic error of the two modes is
4.8\%, when the correlations from particle ID, tracking,
luminosity, and generator are considered.

To estimate the errors in $\BR\cdot \Gamma^{\EE}$, the uncertainties
from the parametrization of the resonances, the phase space factor
due to the intermediate state in $M_{\pp}$ in $Y(4660)$ decays, the
fit range, and the background shape are also considered, in addition
to those in the cross section measurement. 
If a charged structure in $\pi^{\pm}\psp$ exists (cf. Sec.~\ref{VI}), 
it may affect the determination of the resonant parameters. A test fit to
${\pp\psp}$ invariant-mass spectra is tried. Two components of $Y(4360)$ decays are
included in the fit, one decaying to $\pp\psp$ according to three-body 
phase space (50\%) and the other decaying to $\pi^{\pm}Z_c^{\mp}$ (50\%). 
Since the statistical errors of $Z_c$ mass and width are 
large, the mass is fixed to be $4.05~\gevcs$ when the  two-body phase space is calculated.
No $Z_c$ substructure is included in $Y(4660)$ decays. Fit with either $Y(4360)+Y(4660)$
or $Y(4260)+Y(4360)+Y(4660)$  doesnot result in significant change in the 
resonant parameters of the $Y(4360)$ and $Y(4660)$. Since the charged structure is not 
significant (cf. Sec.~\ref{VI}), the effect due to possible existence of the 
$Z_c$ states is not considered.
The factor ${M_i}/{M(\pp\psp)}$ in Eq.~(\ref{bw}) is removed in the fit when
estimating the uncertainties from resonance parametrization. Half of
the difference on each fit result with and without this
factor is taken as the systematic error of resonance
parametrization. In addition, systematic-error contributions
are determined when the fit range is changed from $[4.0,
5.5]~\gevcs$ to $[4.0, 5.3]~\gevcs$ and, separately, the
background shape is changed from a first-order polynomial to a
constant.

All the errors except that from the background estimation are
summarized in Table~\ref{err}. The uncertainties from particle identification,
tracking, luminosity, and generator are common to the two modes.
The total systematic error is calculated to be 4.8\%.

\begin{table}[htbp]
\caption{Relative systematic errors (in \%) in the $\pp\psp$
production cross section measurement.} \label{err}
\begin{center}
\begin{tabular}{c | c c c}
\hline\hline
  Source & $\ppjpsi$ mode & $\MM$ mode & Common \\\hline
 Part ID &  3.3 & 1.4 & 1.4\\
 Tracking & 2.1  & 1.4 & 2.1\\
 $\jpsi$,$\psp$ mass and $\MMS$ & 0.7 & 0.3 & - \\
 Luminosity & 1.4 & 1.4 & 1.4\\
 Generator & 1.9 & 1.9 & 1.9\\
 Trigger & 1.0 & 1.5 & -\\
 Branching fractions & 1.6 & 10.4 & - \\
 MC statistics & 0.1 & 0.1 & -\\
 \hline
 Sum in quadrature & 4.99 & 10.95 & 3.46\\\hline
 Sum of the two modes & \multicolumn{3}{c}{4.8}\\
 \hline\hline
\end{tabular}
\end{center}
\end{table}

\section{Intermediate states}\label{VI}

We search for charged charmoniumlike structures in
both $\psi(2S)$ decay modes of the $\pi^{\pm}\psp$ system
from $Y(4360)$ or $Y(4660)$ decays.
For the $Y(4360)$ subsample, $4.0~\gevcs <
M_{\pp\psp} < 4.5~\gevcs$ is required; for the $Y(4660)$
subsample, $4.5~\gevcs < M_{\pp\psp} < 4.9~\gevcs$ is
required.

\begin{figure*}[htbp]
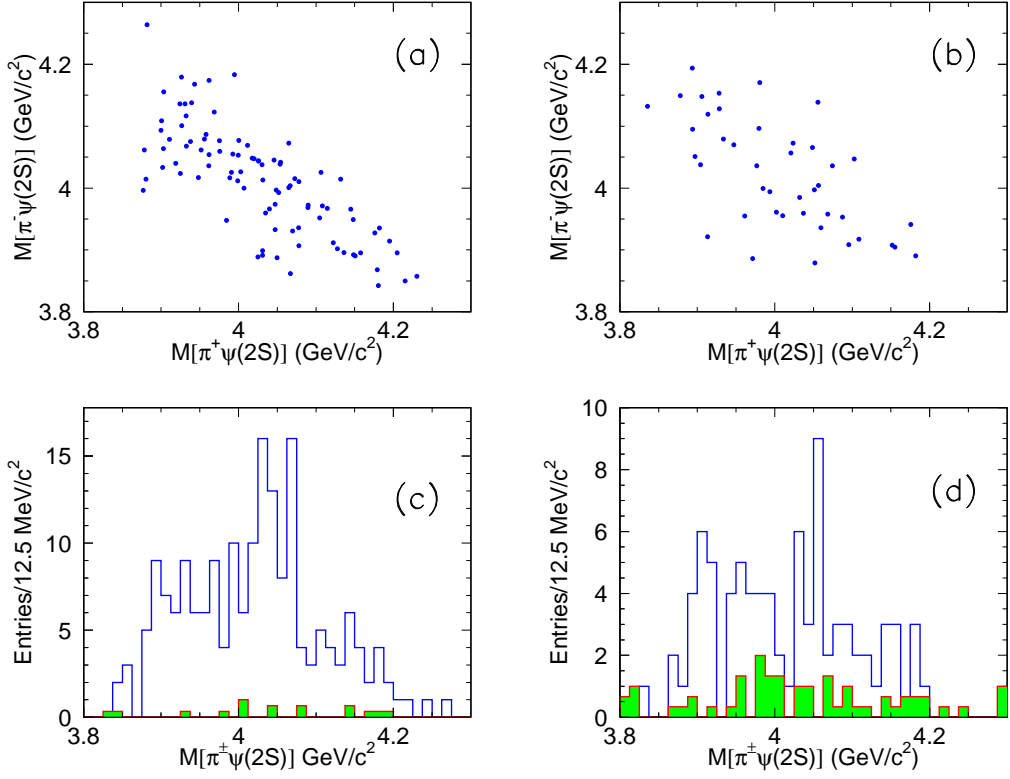

\centering
 \psfig{file=mppsp-y43-ppjpsi-scat-draft.epsi,width=7cm}
 \psfig{file=mppsp-y43-ll-scat-draft.epsi,width=7cm} \\
 \psfig{file=mppsp-ppjpsi-y43-hist-draft.epsi,width=7cm}
 \psfig{file=mppsp-ll-y43-hist-draft.epsi,width=7cm}
\caption{The scatter plots of $M_{\pi^-\psp}$ versus $M_{\pi^+\psp}$
for the $Y(4360)$-subsample events in (a) the
$\ppjpsi$ mode and (b) the $\MM$ mode. Panels (c)
and (d) show the sum of the $M_{\pip\psp}$ and $M_{\pim\psp}$
distributions in the $\ppjpsi$ and $\MM$ modes, respectively. The
shaded histograms are the backgrounds from the normalized $\psp$
mass sidebands.} \label{scat-y43}
\end{figure*}

\begin{figure}[htbp]
 \psfig{file=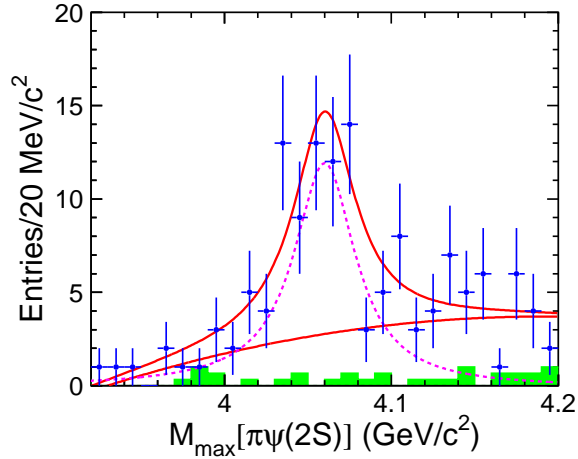,angle=-90,width=7.5cm}
\caption{The distribution of $M_{\rm max}(\pi^{\pm}\psp)$ from
$Y(4360)$-subsample decays. The points with error
bars represent the data; the histogram is from the sidebands
and normalized to the signal region; the solid curve is the
best fit and the dashed curve is the signal parametrized by a
Breit-Wigner function. } \label{mppsp-fit}
\end{figure}

Figure~\ref{scat-y43} shows the scatter plots of $M_{\pim\psp}$
versus $M_{\pip\psp}$ and the one-dimensional projections in
the $Y(4360)$ subsample. There is an excess evident
at around $4.05~\gevcs$ in the $\pi^\pm\psp$ invariant-mass
distributions in both modes. An unbinned maximum-likelihood fit is
performed on the distribution of $M_{\rm max}(\pi^{\pm}\psp)$, the
maximum of $M(\pi^+\psp)$ and $M(\pi^-\psp)$, simultaneously
with both modes. The excess is parameterized with a
Breit-Wigner function and the non-resonant
non-interfering background with a second-order
polynomial function. The fit yields a mass of $(4060\pm 3)~\mevcs$
and a width of $(45\pm 11)~\mev$ for the excess, as shown in
Fig.~\ref{mppsp-fit}. Here, the errors are statistical only.

An MC sample for $Y(4360)\to \pi^\mp+Z^\pm$ and $Z^\pm\to
\pi^\pm+\psp$ is generated to simulate the excess seen in
the data. In the simulation, the mass of $Z^\pm$ is
$4050~\mevcs$ and the width is $40~\mev$. A fit to the
simulated $M_{\rm max}(\pi^\pm\psp)$ distribution yields $M
= (4056\pm 1)~\mevcs$ and $\Gamma = (40.8\pm 2.2)~\mev$. The shift
in the mass is due to the fact that the reflection of the
signal may have a larger $\pi^\pm\psp$ mass than the proper
combination, thus biasing $M_{\rm max}(\pi^{\pm}\psp)$.
We shift the measured mass by $\Delta M = -6~\mevcs$
to account for this effect and assign $1~\mevcs$ as
its systematic error. $\Delta \Gamma = 3.0~\mev$ is taken as the
systematic error of the measured width.

After the bias correction based on MC simulation, we obtain
a mass of $(4054\pm 3({\rm stat.})\pm 1({\rm
syst.}))~\mevcs$ and a width of $(45\pm 11({\rm
stat.})\pm 6({\rm syst.}))~\mev$ for the $Z^\pm$ structure in
the $\pi^{\pm}\psp$ system. The systematic uncertainties
from the parametrization of the resonances, the phase space factor
due to the $J^P$ assignment of the structure, the fit range,
and the background shape are considered. The lowest statistical
significance of the signal is $3.5\sigma$ when comparing
without the Breit-Wigner component.

\begin{figure*}[htbp]
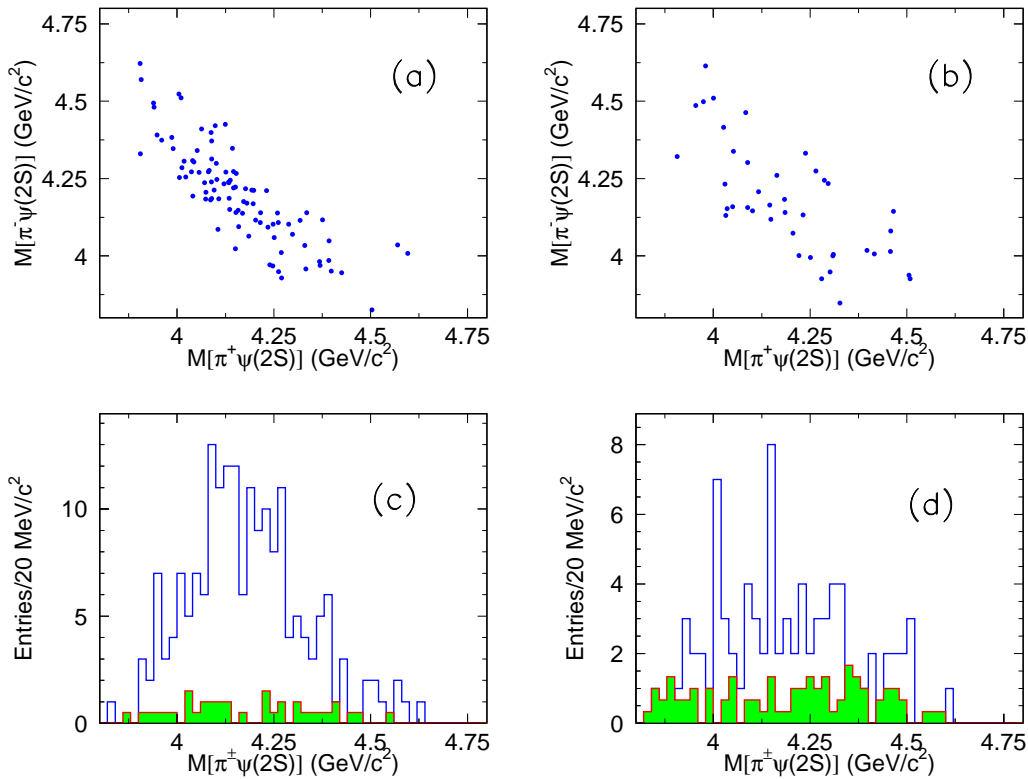

\centering
 \psfig{file=mppsp-y46-ppjpsi-dt-scat-draft.epsi,width=7cm}
 \psfig{file=mppsp-ll-y46-scat-draft.epsi,width=7cm}\\
 \psfig{file=mppsp-dt-y46-ppjpsi-hist-draft.epsi,width=7cm}
 \psfig{file=mppsp-ll-dt-y46-hist-draft.epsi,width=7cm}
\caption{The scatter plots of $M_{\pi^-\psp}$ versus
$M_{\pi^+\psp}$ for the $Y(4660)$ events in the (a) $\ppjpsi$ mode
and (b) the $\MM$ mode. Panels (c) and (d) show the sum of the
$M_{\pip\psp}$ and $M_{\pim\psp}$ distributions in the $\ppjpsi$
and $\MM$ modes, respectively. The shaded histograms are the
backgrounds from the normalized sidebands.} \label{scat-y46}
\end{figure*}

Figure~\ref{scat-y46} shows the scatter plots of $M_{\pim\psp}$
versus $M_{\pip\psp}$ and the one-dimensional projections in
the $Y(4660)$ subsample. This subsample is
limited in statistics---there is no significant structure in
the $\pi^\pm\psp$ system.

\section{Summary}

In summary, the $\EE\to \pp\psp$ cross section is measured from
$4.0$ to $5.5~\gev$ with the full data sample of the Belle
experiment using the ISR technique. The parameters
of the $Y(4360)$ and $Y(4660)$ resonances are
determined; our results agree with and supersede the
previous Belle determination~\cite{pppsp}. Our
results also agree with the BaBar measurement~\cite{babar_pppsp_new}
but with better precision.

We search for a possible charged charmonium-like structure
in $M_{\pi^{\pm}\psp}$ distribution. We find an
excess at $M_{\pi^{\pm}\psp} = 4.05~\gevcs$ in the $Y(4360)$ decays
with a $3.5\sigma$ significance. More data from the
BESIII~\cite{bes3} and the Belle II~\cite{belle2} experiments will
enable a search with improved sensitivity.

\acknowledgments

We thank the KEKB group for the excellent operation of the
accelerator; the KEK cryogenics group for the efficient
operation of the solenoid; and the KEK computer group,
the National Institute of Informatics, and the 
PNNL/EMSL computing group for valuable computing
and SINET4 network support.  We acknowledge support from
the Ministry of Education, Culture, Sports, Science, and
Technology (MEXT) of Japan, the Japan Society for the 
Promotion of Science (JSPS), and the Tau-Lepton Physics 
Research Center of Nagoya University; 
the Australian Research Council and the Australian 
Department of Industry, Innovation, Science and Research;
Austrian Science Fund under Grant No.~P 22742-N16 and P 26794-N20;
the National Natural Science Foundation of China under Contracts 
No.~10575109, No.~10775142, No.~10875115, No.~11175187, and  No.~11475187;
the Chinese Academy of Science Center for Excellence in Particle Physics; 
the Ministry of Education, Youth and Sports of the Czech
Republic under Contract No.~LG14034;
the Carl Zeiss Foundation, the Deutsche Forschungsgemeinschaft
and the VolkswagenStiftung;
the Department of Science and Technology of India; 
the Istituto Nazionale di Fisica Nucleare of Italy; 
National Research Foundation (NRF) of Korea Grants
No.~2011-0029457, No.~2012-0008143, No.~2012R1A1A2008330, 
No.~2013R1A1A3007772, No.~2014R1A2A2A01005286, No.~2014R1A2A2A01002734, 
No.~2014R1A1A2006456;
the Basic Research Lab program under NRF Grant No.~KRF-2011-0020333, 
No.~KRF-2011-0021196, Center for Korean J-PARC Users, No.~NRF-2013K1A3A7A06056592; 
the Brain Korea 21-Plus program and the Global Science Experimental Data 
Hub Center of the Korea Institute of Science and Technology Information;
the Polish Ministry of Science and Higher Education and 
the National Science Center;
the Ministry of Education and Science of the Russian Federation and
the Russian Foundation for Basic Research;
the Slovenian Research Agency;
the Basque Foundation for Science (IKERBASQUE) and 
the Euskal Herriko Unibertsitatea (UPV/EHU) under program UFI 11/55 (Spain);
the Swiss National Science Foundation; the National Science Council
and the Ministry of Education of Taiwan; and the U.S.\
Department of Energy and the National Science Foundation.
This work is supported by a Grant-in-Aid from MEXT for 
Science Research in a Priority Area (``New Development of 
Flavor Physics'') and from JSPS for Creative Scientific 
Research (``Evolution of Tau-lepton Physics'').

\newpage

\appendix

\section{Fits to $\EE\to\pp\psp$ using $\pp\jpsi$ mode only} \label{App:Appendix0}

\begin{figure*}[htbp]
 \psfig{file=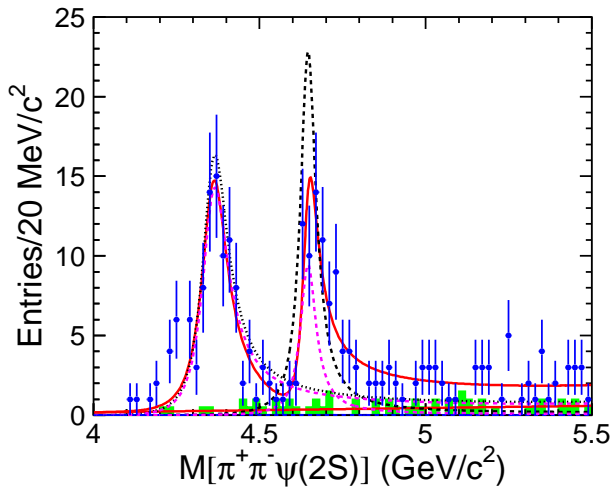, height=8.cm, angle=-90}
\caption{The $\pp\psp$ invariant-mass distributions and the fit
results for the $\ppjpsi$ mode only. The details of the plot are
the same as those in Fig.~\ref{fit}.} \label{fit-ppjpsi}
\end{figure*}

To compare with the previous measurement from Belle~\cite{pppsp},
a fit to the $\pp\jpsi$ mode only is also performed; the fit
results are shown in Fig.~\ref{fit-ppjpsi} and
Table~\ref{tab-ppjpsi}. There are differences in the fit results
between this measurement and the previous one~\cite{pppsp}; this
can be explained by the strong correlation between the parameters.
For example, the correlation coefficient between $M_{Y(4660)}$ and
$\phi$ is 0.86 for one solution (or $-0.76$ for the other solution)
in the fit shown in Tables~\ref{two_sol}~and~\ref{corr}.

\begin{table}
\caption{Results of the fits to the $\pp\psp$ invariant-mass
spectra, using the $\ppjpsi$ mode only. The details are the same
as those in Table~\ref{two_sol}.} \label{tab-ppjpsi}
\begin{center}
\begin{tabular}{c |c c }
\hline\hline
     Parameters     & ~~Solution~I~~~ & ~~~Solution~II~~  \\\hline
 $M_{Y(4360)}$        & \multicolumn{2}{c}{$4358\pm6\pm2$} \\
 $\Gamma_{Y(4360)}$   & \multicolumn{2}{c}{$96\pm10\pm6$}  \\
 $\BR[Y(4360)\to\pp\psp]\cdot\Gamma_{Y(4360)}^{\EE}$
                    & $9.4\pm0.8\pm0.7$ & $10.8\pm0.7\pm0.7$   \\
 $M_{Y(4660)}$       & \multicolumn{2}{c}{$4644\pm7\pm5$} \\
 $\Gamma_{Y(4660)}$  & \multicolumn{2}{c}{$57\pm9\pm5$}   \\
 $\BR[Y(4660)\to\pp\psp]\cdot\Gamma_{Y(4660)}^{\EE}$
                    & $3.1\pm0.5\pm0.4$ & $7.6\pm1.3\pm0.9$   \\
 $\phi$             & $10\pm17\pm12$ & $288\pm10\pm5$  \\
 \hline\hline
\end{tabular}
\end{center}
\end{table}

\begin{figure*}[htbp]
 \psfig{file=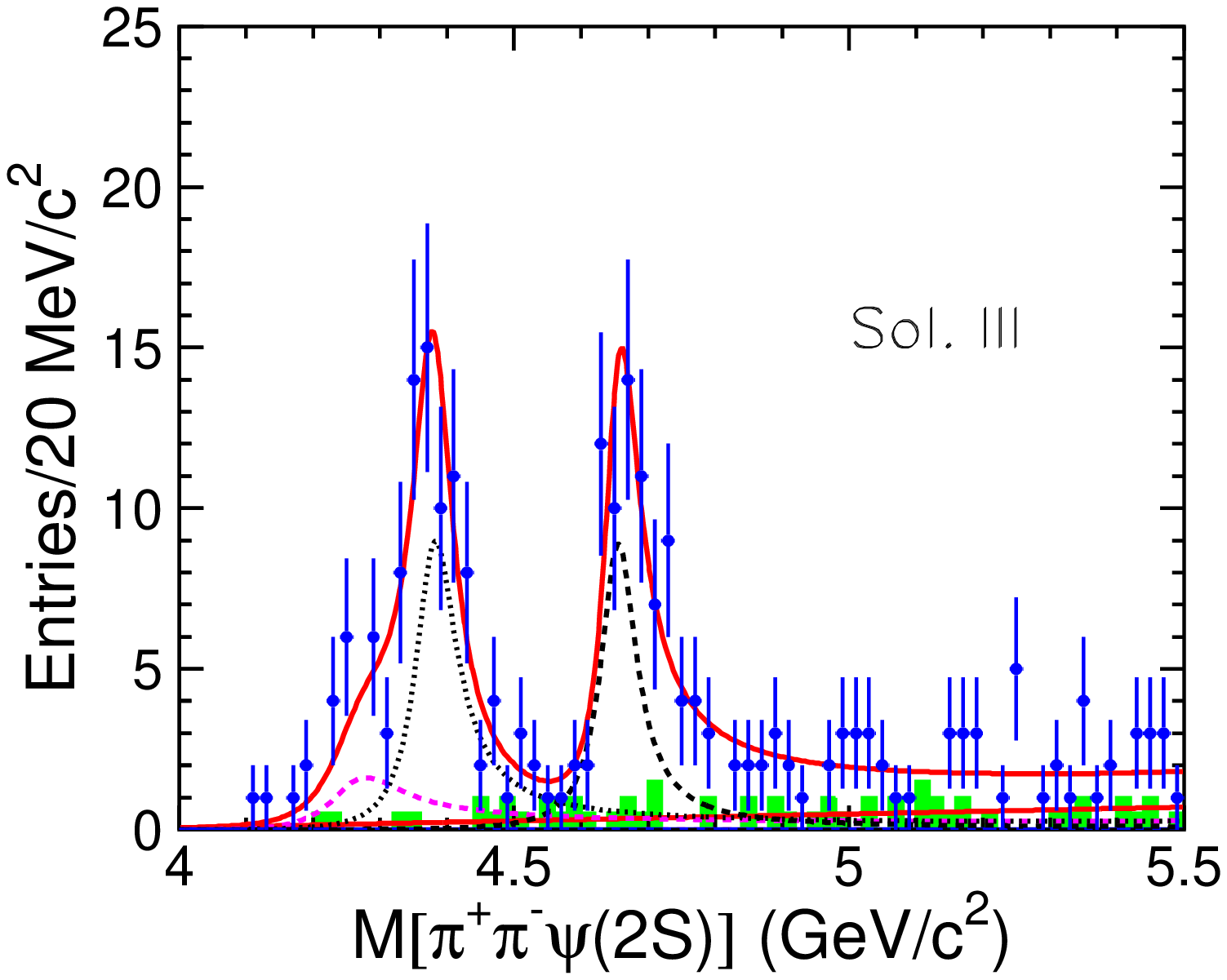, height=8.cm, angle=-90}
 \psfig{file=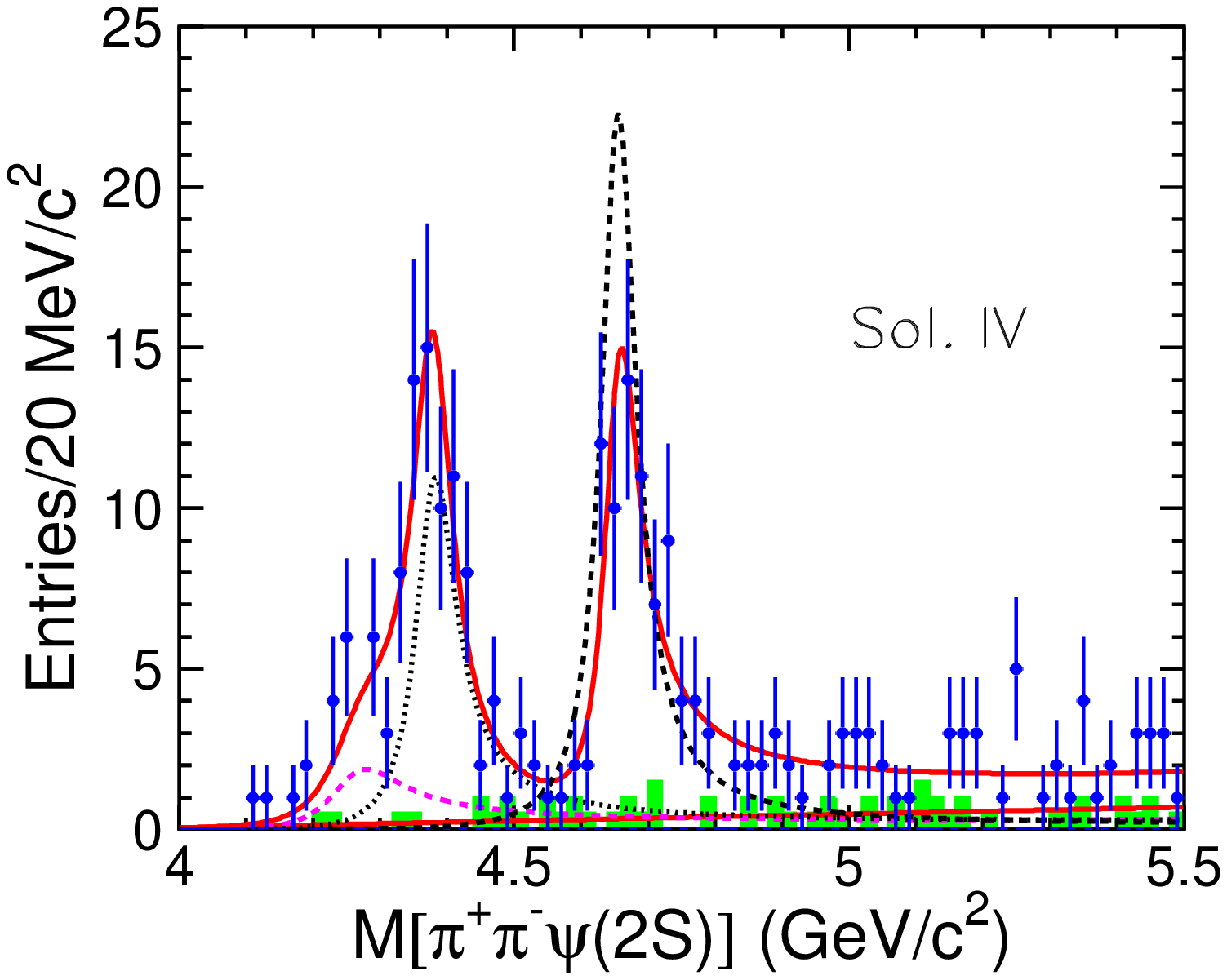, height=8.cm, angle=-90}\\
 \psfig{file=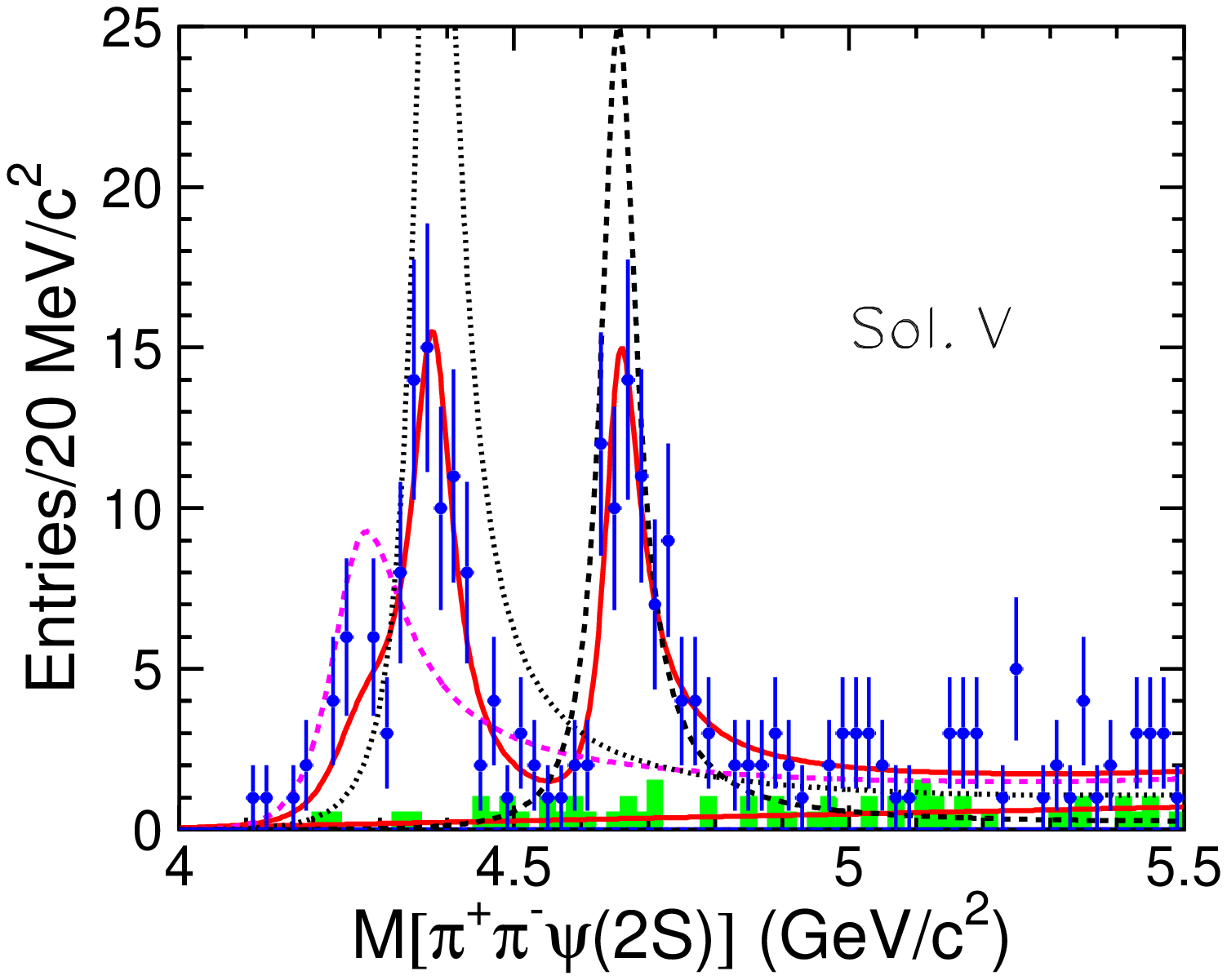, height=8.cm, angle=-90}
 \psfig{file=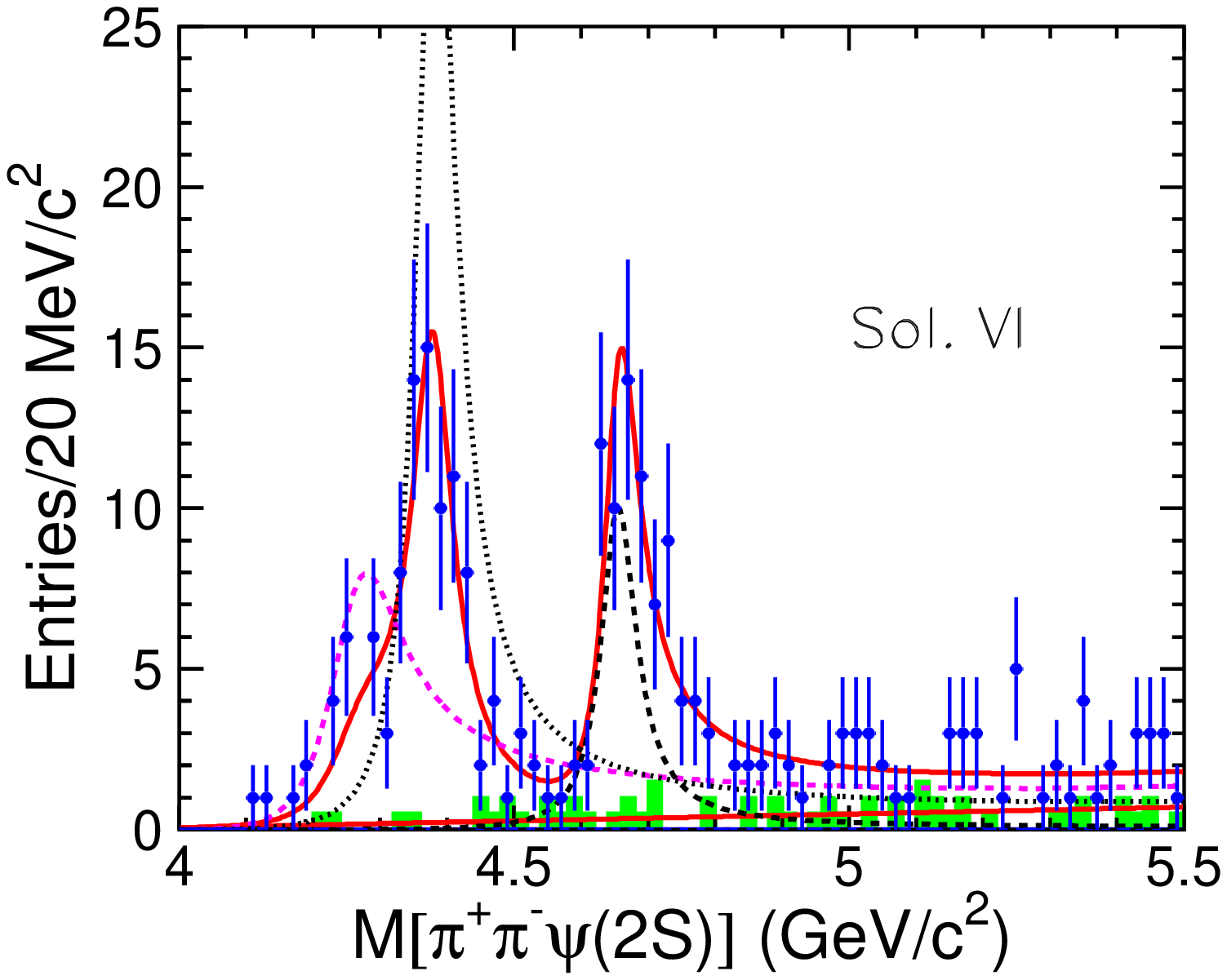, height=8.cm, angle=-90}
\caption{The four solutions from the fit to the $\pp\psp$
invariant-mass spectra with the $Y(4260)$ included but for the
$\ppjpsi$ mode only. The details are the same as those in
Fig.~\ref{fit-3r}.} \label{fit-ppjpsi-3r}
\end{figure*}

The fit to the $\pp\jpsi$ mode only with the coherent sum of
$Y(4260)$, $Y(4360)$, and $Y(4660)$ is shown in
Fig.~\ref{fit-ppjpsi-3r} and Table~\ref{tab-ppjpsi-3r}. The
statistical significance of the $Y(4260)$ is 
$2.8\sigma$ in this fit.

\begin{table*}
\caption{Results of the fits to the $\pp\psp$ invariant-mass
spectra in $\ppjpsi$ mode only using three resonances, the
$Y(4260)$, $Y(4360)$ and $Y(4660)$. The details are the same as
those from Table~\ref{three_res}.} \label{tab-ppjpsi-3r}
\begin{center}
\begin{tabular}{c |c c c c}
\hline\hline
     Parameters     & ~~Solution~III~~~ & ~~~Solution~IV~~  & ~~Solution~V~~ & ~~~Solution~VI~~\\\hline
 $M_{Y(4260)}$        & \multicolumn{4}{c}{4259~(fixed)} \\
 $\Gamma_{Y(4260)}$   & \multicolumn{4}{c}{134~(fixed)}  \\
 $\BR[Y(4260)\to\pp\psp]\cdot\Gamma_{Y(4260)}^{\EE}$
                    & $1.6\pm0.6\pm0.4$ & $1.8\pm0.8\pm0.6$ & $9.1\pm1.2\pm0.7$ & $7.8\pm 1.1\pm0.8$  \\
 $M_{Y(4360)}$        & \multicolumn{4}{c}{$4378\pm9\pm6$} \\
 $\Gamma_{Y(4360)}$   & \multicolumn{4}{c}{$74\pm14\pm3$}  \\
 $\BR[Y(4360)\to\pp\psp]\cdot\Gamma_{Y(4360)}^{\EE}$
                    & $4.5\pm1.0\pm0.4$ & $5.5\pm1.4\pm0.6$  & $19.1\pm 2.8\pm1.1$ & $15.7\pm 2.3\pm1.6$\\
 $M_{Y(4660)}$       & \multicolumn{4}{c}{$4654\pm7\pm6$} \\
 $\Gamma_{Y(4660)}$  & \multicolumn{4}{c}{$65\pm10\pm3$}   \\
 $\BR[Y(4660)\to\pp\psp]\cdot\Gamma_{Y(4660)}^{\EE}$
                    & $3.3\pm0.6\pm0.3$ & $8.3\pm1.0\pm0.9$ & $9.3\pm 1.2\pm1.2$ & $3.7\pm 0.7\pm0.5$\\
 $\phi_1$             & $282\pm25\pm24$ & $270\pm27\pm28$ & $130\pm 5\pm3 $ & $142\pm 6\pm7$\\
 $\phi_2$             & $359\pm19\pm3$ & $243\pm17\pm20$ & $337\pm 10\pm7$ & $93\pm 25\pm17$ \\
\hline\hline
\end{tabular}
\end{center}
\end{table*}

\section{Cross section of $\EE\to\pp\psp$} \label{App:AppendixA}

\begin{table}[htbp]
\caption{Measured $\EE\to\pp\psp$ cross section for center of mass
energy ($E_{\rm cm}$) from $4.0~\gevcs$ to $5.5~\gevcs$. The
errors are the sums of statistical errors of signal and background
events and the systematic errors. } \label{xs-points}
\begin{center}
\begin{tabular}{c r @{$\pm$} l| c r @{$\pm$} l | c r @{$\pm$} l}
\hline\hline
  $E_{\rm cm}$ ($\gev$) & \multicolumn{2}{c}{Cross section ($\rm pb$)} & $E_{\rm cm}$ ($\gev$) 
& \multicolumn{2}{c}{Cross section ($\rm pb$)} & $E_{\rm cm}$ ($\gev$) & \multicolumn{2}{c}{Cross section ($\rm pb$)}  \\\hline
   4.01 &  $ -19.1 $&$  31.4$ &    4.51  & $ 18.4  $&$   8.4$ &    5.01 &  $  3.7 $&$    4.6$ \\
   4.03 &  $ -15.3 $&$  24.9$ &    4.53  & $ 14.9  $&$   7.6$ &    5.03 &  $ 10.1 $&$    5.3$ \\
   4.05 &  $ -12.8 $&$  20.7$ &    4.55  & $  5.5  $&$   5.2$ &    5.05 &  $  5.8 $&$    4.3$ \\
   4.07 &  $ -11.0 $&$  17.8$ &    4.57  & $ -0.5  $&$   4.6$ &    5.07 &  $  3.6 $&$    3.6$ \\
   4.09 &  $ -9.7  $&$  15.7$ &    4.59  & $  8.2  $&$   5.8$ &    5.09 &  $  3.5 $&$    3.6$ \\
   4.11 &  $  -0.5 $&$  12.9$ &    4.61  & $ 10.9  $&$   6.4$ &    5.11 &  $ -2.6 $&$    3.5$ \\
   4.13 &  $ -0.5  $&$  11.7$ &    4.63  & $ 33.0  $&$  10.2$ &    5.13 &  $ -2.6 $&$    3.4$ \\
   4.15 &  $ -0.5  $&$  10.7$ &    4.65  & $ 29.7  $&$   9.6$ &    5.15 &  $  7.4 $&$    4.5$ \\
   4.17 &  $  5.9  $&$   9.0$ &    4.67  & $ 37.3  $&$  10.6$ &    5.17 &  $  7.3 $&$    4.4$ \\
   4.19 &  $  5.4  $&$  11.0$ &    4.69  & $ 34.1  $&$  10.1$ &    5.19 &  $  5.2 $&$    3.9$ \\
   4.21 &  $ -6.0  $&$   9.4$ &    4.71  & $ 20.4  $&$   7.9$ &    5.21 &  $ -0.6 $&$    3.0$ \\
   4.23 &  $ 20.4  $&$  11.7$ &    4.73  & $ 22.7  $&$   8.2$ &    5.23 &  $  5.0 $&$    3.8$ \\
   4.25 &  $ 29.2  $&$  13.1$ &    4.75  & $  9.6  $&$   5.7$ &    5.25 &  $ 10.5 $&$    5.0$ \\
   4.27 &  $ 8.9   $&$   9.9$ &    4.77  & $ 14.5  $&$   6.7$ &    5.27 &  $ -0.6 $&$    2.9$ \\
   4.29 &  $ 26.5  $&$  11.9$ &    4.79  & $  9.4  $&$   5.6$ &    5.29 &  $  1.2 $&$    2.6$ \\
   4.31 &  $ 25.3  $&$  11.4$ &    4.81  & $  4.3  $&$   5.2$ &    5.31 &  $  6.5 $&$    4.0$ \\
   4.33 &  $ 61.5  $&$  16.5$ &    4.83  & $  1.9  $&$   4.5$ &    5.33 &  $ -0.6 $&$    2.7$ \\
   4.35 &  $ 67.1  $&$  16.9$ &    4.85  & $  6.6  $&$   4.8$ &    5.35 &  $  4.6 $&$    4.0$ \\
   4.37 &  $ 80.1  $&$  18.0$ &    4.87  & $  6.5  $&$   4.7$ &    5.37 &  $ -0.6 $&$    2.6$ \\
   4.39 &  $ 40.4  $&$  12.9$ &    4.89  & $ 11.1  $&$   5.7$ &    5.39 &  $  2.7 $&$    2.9$ \\
   4.41 &  $ 42.7  $&$  13.0$ &    4.91  & $  6.3  $&$   4.6$ &    5.41 &  $ -0.6 $&$    2.6$ \\
   4.43 &  $ 31.0  $&$  11.0$ &    4.93  & $  4.0  $&$   4.0$ &    5.43 &  $  2.6 $&$    3.4$ \\
   4.45 &  $  9.7  $&$   6.8$ &    4.95  & $ -2.8  $&$   3.8$ &    5.45 &  $  4.1 $&$    3.2$ \\
   4.47 &  $ 12.8  $&$   7.4$ &    4.97  & $  1.6  $&$   4.1$ &    5.47 &  $  5.6 $&$    3.5$ \\
   4.49 &  $  6.0  $&$   5.6$ &    4.99  & $  6.0  $&$   4.4$ &    5.49 &  $ -0.6 $&$    2.4$ \\ \hline

\end{tabular}
\end{center}
\end{table}


\begin{thebibliography}{**}

\bibitem{review} For recent reviews, see N.~Brambilla {\it et al.},
\Journal\EPJC{71}{1534}{2011}; N.~Brambilla {\it et al.},
arXiv:1404.3723v2.
\bibitem{ycz_review}
  C.~-Z.~Yuan,
\Journal\IJMP{A29}{1430046}{2014}.
\bibitem{babay4260} B.~Aubert {\it et al.} (BaBar Collaboration),
\Journal\PRL{95}{142001}{2005}.
\bibitem{cleoy} Q. He {\it et al.} (CLEO Collaboration),
\Journal\PRD{74}{091104(R)}{2006}.
\bibitem{belley} C.~Z.~Yuan {\it et al.} (Belle Collaboration),
\Journal\PRL{99}{182004}{2007}.
\bibitem{babay4324} B.~Aubert {\it et al.} (BaBar Collaboration),
\Journal\PRL{98}{212001}{2007}.
\bibitem{pppsp} X.~L. Wang {\it et al.} (Belle Collaboration),
\Journal\PRL{99}{142002}{2007}.
\bibitem{babary_new} J.~P.~Lees {\it et al.} (BaBar Collaboration),
\Journal\PRD{86}{051102(R)}{2012}.
\bibitem{belley_new} Z.~Q.~Liu {\it et al.} (Belle
Collaboration), \Journal\PRL{110}{252002}{2013}.
\bibitem{babar_pppsp_new} B. Aubert {\it et al.} (BaBar Collaboration),
\Journal\PRD{89}{111103}{2014}.
\bibitem{etajpsi}X.~L. Wang {\it et al.} (Belle Collaboration),
\Journal\PRD{87}{051101(R)}{2013}.
\bibitem{zc3900} M.~Ablikim {\it et al.} (BESIII Collaboration),
\Journal\PRL{110}{252001}{2013}.
\bibitem{zc4020} M.~Ablikim {\it et al.} (BESIII Collaboration),
\Journal\PRL{111}{242001}{2013}.
\bibitem{Belle} A.~Abashian {\it et al.} (Belle Collaboration),
\Journal\NIMA{479}{117}{2002}; also see detector section in J. Brodzicka {\it et al.},
\Journal\PTEP{}{04D001}{2012}.
\bibitem{KEKB} S.~Kurokawa and E.~Kikutani,
\Journal\NIMA{499}{1}{2003}
and other papers included in this volume; T.~Abe {\it et al.},
\Journal\PTEP{}{03A001}{2013} and following articles up to 03A011.
\bibitem{geant3} R.~Brun {\it et al.}, 
GEANT 3.21, CERN DD/EE/84-1, 1984.
\bibitem{phokhara} G.~Rodrigo {\it et al.},
\Journal\EPJC{24}{71}{2002}. For a review on the generator, see:
S.~Actis {\it et al.}, \Journal\EPJC{66}{585}{2010}.
\bibitem{pid} E.~Nakano,
\Journal\NIMA{494}{402}{2002}.
\bibitem{EID} K.~Hanagaki {\it et al.},
\Journal\NIMA{485}{490}{2002}.
\bibitem{MUID} A.~Abashian {\it et al.},
\Journal\NIMA{491}{69}{2002}.
\bibitem{PDG} J.~Beringer {\it et al.} (Particle Data Group),
\Journal\PRD{86}{010001}{2012}.
\bibitem{mppjpsi}
$M_{\pp\jpsi} = M_{\pp\LL}-M_{\LL}+m_{\jpsi}$ is used to cancel the
lepton-pair mass resolution in the $\pp\jpsi$ invariant mass
spectrum; here, $m_{\jpsi}$ is the nominal mass of
$\jpsi$~\cite{PDG}.
\bibitem{mpppsp}
$M_{\pp\psp} = M_{\pp\pp\LL} - M_{\pp\LL}+ m_{\psp} $ for the
$\ppjpsi$ mode and $M_{\pp\psp} = M_{\pp\MM} - M_{\MM} + m_{\psp}$
for the $\MM$ mode, where $m_{\psp}$ is the $\psp$ norminal
mass~\cite{PDG}.
\bibitem{kuraev} E.~A.~Kuraev and V.~S.~Fadin,
Sov. J. Nucl. Phys.  {\bf 41}, 466 (1985) [Yad. Fiz. {\bf 41}, 733
(1985)].

\bibitem{bes3}M.~Ablikim {\it et al.} (BESIII Collaboration),
\Journal\NIMA{614}{345}{2010}.
\bibitem{belle2}T.~Abe {\it et al.} (Belle II Collaboration),
arXiv:1011.0352.
\end{thebibliography}
\end{document}